\documentclass[longauth]{aa}
\usepackage[utf8]{inputenc}

\usepackage{amsmath}
\usepackage{txfonts}
\usepackage{graphicx}
\usepackage{units}
\usepackage[breaklinks, colorlinks, citecolor=blue, linkcolor=blue]{hyperref}

\usepackage{siunitx}
\usepackage{natbib}
\usepackage{blindtext}
\usepackage{longtable}
\usepackage{multirow, bigdelim}
\usepackage{pdflscape}

\let\orgautoref\autoref
\renewcommand{\autoref}
        {\def\equationautorefname{Eq.}%
         \def\figureautorefname{Fig.}%
         \def\sectionautorefname{Sect.}%
         \def\subsectionautorefname{Sect.}%
         \def\subsubsectionautorefname{Sect.}%
         \orgautoref}

\newcommand*\samethanks[1][\value{footnote}]{\footnotemark[#1]}

\newcommand{\tx}[1]{\mathrm{#1}}

\newcommand{\fourteenday}{14.3-day\space}
\newcommand{\fifteenday}{15.0-day\space}
\newcommand{\twoday}{2.62-day\space}
\newcommand{\oneday}{1.62-day\space}
\newcommand{\Ptransiting}{{\SI{2.6162745\pm0.0000030}{\day}}}
\newcommand{\Rtransiting}{{\SI{1.150\pm0.040}{R_\oplus}}}
\newcommand{\Mtransiting}{{\SI{1.21\pm0.42}{M_\oplus}}}
\newcommand{\rhotransiting}{{\SI{4.4\pm1.6}{\gram\per\centi\meter\cubed}}}
\newcommand{\Prv}{{\SI{14.303\pm0.035}{\day}}}
\newcommand{\Mrv}{{\SI{5.27\pm0.74}{M_\oplus}}}

\makeatletter

\def\instrefs#1{{\def\scsep{\def\scsep{,}}\@for\w:=#1\do{\scsep\ref{inst:\w}}}}
\renewcommand{\inst}[1]{\unskip$^{\instrefs{#1}}$}

\DeclareSIUnit\parsec{pc}
\DeclareSIUnit\arcsecond{\arcsec}
\DeclareSIUnit\arcminute{^\prime}
\DeclareSIUnit\pixel{pix}
\DeclareSIUnit\year{yr}
\DeclareSIUnit\Np{Np}
\usepackage{color}

\begin{document}

\title{Discovery and mass measurement of the hot, transiting, Earth-sized planet, GJ~3929~b\thanks{RV data and stellar activity indices are only available in electronic form
        at the CDS via anonymous ftp to cdsarc.u-strasbg.fr (130.79.128.5)
        or via \url{http://cdsweb.u-strasbg.fr/cgi-bin/qcat?J/A+A/}}}
\titlerunning{Discovery of a planetary system around {GJ~3929}}

\author{J.~Kemmer\inst{lsw}\thanks{Fellow of the International Max Planck Research School for Astronomy and Cosmic Physics at the University of Heidelberg (IMPRS-HD).}
\and S.~Dreizler\inst{iag}
\and D.~Kossakowski\inst{mpia}
\and S.~Stock\inst{lsw}
\and A.~Quirrenbach\inst{lsw}
\and J.\,A.~Caballero\inst{cabesac}
\and P.\,J.~Amado\inst{iaa}
\and K.\,A.~Collins\inst{cfa}
\and N.~Espinoza\inst{stsci}
\and E.~Herrero\inst{ice,ieec}
\and J.\,M.~Jenkins\inst{ames}
\and D.\,W.~Latham\inst{cfa}
\and J.~Lillo-Box\inst{cabesac}
\and N.~Narita\inst{komaba,aco,iac}
\and E.~Pall\'e\inst{iac,ull}
\and A.~Reiners\inst{iag}
\and I.~Ribas\inst{ice,ieec}
\and G.~Ricker\inst{mit}
\and E.~Rodr\'iguez\inst{iaa}
\and S.~Seager\inst{mit,mit_planetarysciences,mit_aeronautics}
\and R.~Vanderspek\inst{mit}
\and R.~Wells\inst{unibe}
\and J.~Winn\inst{princeton}
\and {F.\,J.~Aceituno}\inst{iaa}
\and V.\,J.\,S.~B\'ejar\inst{iac,ull}
\and T.~Barclay\inst{goddard,umb}
\and P.~Bluhm\inst{lsw}\samethanks
\and P.~Chaturvedi\inst{tls}
\and C.~Cifuentes\inst{cabesac}
\and K.\,I.~Collins\inst{gmu}
\and M.~Cort\'es-Contreras\inst{cabesac}
\and B.-O.~Demory\inst{unibe}
\and M.\,M.~Fausnau{g}h\inst{mit}
\and A.~Fukui\inst{komaba,iac}
\and Y.~G\'omez~Maqueo~Chew\inst{unam}
\and D.~Galad\'i-Enr\'iquez\inst{caha}
\and T.~Gan\inst{tsu}
\and M.~Gillon\inst{liege}
\and A.~Golovin\inst{lsw}\samethanks
\and A.\,P.~Hatzes\inst{tls}
\and Th.~Henning\inst{mpia}
\and C.~Huang\inst{mit}
\and S.\,V.~Jeffers\inst{mig}
\and A.~Kaminski\inst{lsw}
\and M.~Kunimoto\inst{mit}
\and M.~Kürster\inst{mpia}
\and M.\,J.~L\'opez-Gonz\'alez\inst{iaa}
\and M.~Lafarga\inst{ice,ieec,warwick}
\and R.~Luque\inst{iaa}
\and J.~McCormac\inst{warwick}
\and K.~Molaverdikhani\inst{lmu,origins,lsw}
\and D.~Montes\inst{ucm}
\and J.\,C.~Morales\inst{ice,ieec}
\and V.\,M.~Passegger\inst{ham,homer}
\and S.~Reffert\inst{lsw}
\and L.~Sabin\inst{uname}
\and P.~Sch\"ofer\inst{iag}
\and N.~Schanche\inst{unibe}
\and M.~Schlecker\inst{mpia}
\and U.~Schroffenegger\inst{unibe}
\and R.\,P.~Schwarz\inst{pata}
\and A.~Schweitzer\inst{ham}
\and {A.~Sota}\inst{iaa}
\and P.~Tenenbaum\inst{seti,ames}
\and T.~Trifonov\inst{mpia}
\and S.~Vanaverbeke\inst{leuven,iris}
\and M.~Zechmeister\inst{iag}
}

\institute{
\label{inst:lsw}Landessternwarte, Zentrum f\"ur Astronomie der Universit\"at Heidelberg, K\"onigstuhl 12, 69117 Heidelberg, Germany
\and \label{inst:iag}Institut f\"ur Astrophysik, Georg-August-Universit\"at, Friedrich-Hund-Platz 1, 37077 G\"ottingen, Germany
\and \label{inst:mpia}Max-Planck-Institut f\"{u}r Astronomie, K\"{o}nigstuhl  17, 69117 Heidelberg, Germany
\and \label{inst:cabesac}Centro de Astrobiolog\'ia (CSIC-INTA), ESAC, Camino bajo del castillo s/n, 28692 Villanueva de la Ca\~nada, Madrid, Spain
\and \label{inst:iaa}Instituto de Astrof\'isica de Andaluc\'ia (CSIC), Glorieta de la Astronom\'ia s/n, 18008 Granada, Spain
\and \label{inst:cfa}Center for Astrophysics \textbar \ Harvard \& Smithsonian, 60 Garden Street, Cambridge, MA 02138, United States of America
\and \label{inst:stsci}Space Telescope Science Institute, Baltimore, MD 21218, United States of America
\and \label{inst:ice}Institut de Ci\`encies de l'Espai (ICE, CSIC), Campus UAB, c/ de Can Magrans s/n, 08193 Cerdanyola del Vall\`es, Spain
\and \label{inst:ieec}Institut d'Estudis Espacials de Catalunya (IEEC), c/ Gran Capit\`a 2-4, 08034 Barcelona, Spain
\and \label{inst:ames}NASA Ames Research Center, Moffett Field, CA 94035, United States of America
\and \label{inst:komaba}Komaba Institute for Science, The University of Tokyo, 3-8-1 Komaba, Meguro, Tokyo 153-8902, Japan
\and \label{inst:aco}Astrobiology Center, 2-21-1 Osawa, Mitaka, Tokyo 181-8588, Japan
\and \label{inst:iac}Instituto de Astrof\'isica de Canarias (IAC), 38205 La Laguna, Te\-ne\-ri\-fe, Spain
\and \label{inst:ull}Departamento de Astrof\'isica, Universidad de La Laguna, 38206 La Laguna, Tenerife, Spain
\and \label{inst:mit}Department of Physics and Kavli Institute for Astrophysics and Space Research, Massachusetts Institute of Technology, Cambridge, MA 02139, United States of America
\and \label{inst:mit_planetarysciences}Department of Earth, Atmospheric and Planetary Sciences, Massachusetts Institute of Technology, Cambridge, MA 02139, United States of America
\and \label{inst:mit_aeronautics}Department of Aeronautics and Astronautics, Massachusetts Institute of Technology, 77 Massachusetts Avenue, Cambridge, MA 02139, United States of America
\and \label{inst:unibe}Center for Space and Habitability, University of Bern, Gesellschaftsstrasse 6, 3012, Bern, Switzerland
\and \label{inst:princeton}Department of Astrophysical Sciences, Princeton University, 4 Ivy Lane, Princeton, NJ 08544, United States of America
\and \label{inst:goddard}NASA Goddard Space Flight Center, 8800 Greenbelt Road, Greenbelt, MD 20771, United States of America
\and \label{inst:umb}University of Maryland, Baltimore County, 1000 Hilltop Circle, Baltimore, MD 21250, United States of America
\and \label{inst:tls}Th\"uringer Landessternwarte Tautenburg, Sternwarte 5, 07778 Tautenburg, Germany
\and \label{inst:gmu}George Mason University, 4400 University Drive, Fairfax, VA 22030, United States of America
\and \label{inst:unam}Universidad Nacional Aut\'onoma de M\'exico, Instituto de Astronom\'ia, AP 70-264, CDMX  04510, M\'exico
\and \label{inst:caha}Centro Astron\'onomico Hispano Alem\'an, Observatorio de Calar Alto, Sierra de los Filabres, E-04550 G\'ergal, Spain
\and \label{inst:tsu}Department of Astronomy and Tsinghua Centre for Astrophysics, Tsinghua University, Beijing 100084, China
\and \label{inst:liege}Astrobiology Research Unit, Universit\'e de Li\`ege, All\'ee du 6 Ao\^ut 19C, B-4000 Li\`ege, Belgium
\and \label{inst:mig}Max-Planck-Institut f\"ur Sonnensystemforschung, Justus-von-Liebig Weg 3, 37077 G\"ottingen, Germany
\and \label{inst:warwick}Department of Physics, University of Warwick, Gibbet Hill Road, Coventry CV4 7AL, United Kingdom
\and \label{inst:lmu}Universit\"ats-Sternwarte, Ludwig-Maximilians-Universit\"at M\"unchen, Scheinerstrasse 1, 81679 M\"unchen, Germany
\and \label{inst:origins}Exzellenzcluster Origins, Boltzmannstraße 2, 85748 Garching, Germany
\and \label{inst:ucm}Departamento de F{\'i}sica de la Tierra y Astrof{\'i}sica \& IPARCOS-UCM (Instituto de F\'{i}sica de Part\'{i}culas y del Cosmos de la UCM), Facultad de Ciencias F{\'i}sicas, Universidad Complutense de Madrid, E-28040 Madrid, Spain
\and \label{inst:ham}Hamburger Sternwarte, Gojenbergsweg 112, 21029 Hamburg, Germany
\and \label{inst:homer}{Homer L. Dodge Department of Physics and Astronomy, University of Oklahoma, 440 West Brooks Street, Norman, OK 73019, United States of America}
\and \label{inst:uname}Universidad Nacional Aut\'onoma de M\'exico, Instituto de Astronom\'ia, AP 106, Ensenada 22800, BC, M\'exico
\and \label{inst:pata}Patashnick Voorheesville Observatory, Voorheesville, NY 12186, United States of America
\and \label{inst:seti}SETI Institute, Mountain View, CA 94043, United States of America
\and \label{inst:leuven}Vereniging Voor Sterrenkunde, Brugge, Belgium \& Centre for mathematical Plasma-Astrophysics, Department of Mathematics, Katholieke Universiteit Leuven, Celestijnenlaan 200B, 3001 Heverlee, Belgium
\and \label{inst:iris}AstroLAB IRIS, Provinciaal Domein ``De Palingbeek'', Verbrandemolenstraat 5, 8902 Zillebeke, Ieper, Belgium
}

\date{Received {12} November 2021 /
    Accepted 14 January 2022}

\abstract
{We report the discovery of GJ~3929~b, a hot Earth-sized planet orbiting the {nearby} M3.5\,V dwarf star, \object{GJ~3929} (G~180--18, TOI-2013). Joint modelling of photometric observations from TESS sectors 24 and 25 together with 73 spectroscopic observations from CARMENES and follow-up transit observations from SAINT-EX{, LCOGT, and OSN} {yields} a {planet} radius of $R_\text{{b}} = \Rtransiting$, a mass of $M_\text{{b}} = \Mtransiting${,} {and an orbital period of} $P_\text{{b}} = \Ptransiting$. {The resulting density of $\rho_\text{{b}}= \rhotransiting$ is compatible with the Earth's mean density of {about} \SI{5.5}{\gram\per\centi\meter\cubed}}. Due to the apparent brightness of the host star {($J=\SI{8.7}{mag}$)} and its small size{,} GJ~3929~b is a {promising} target for atmospheric {characterisation} with the {JWST}.
Additionally, the radial velocity data show evidence for {another planet candidate} with $P_\text{[c]} = \Prv$, which {is likely} unrelated to the stellar rotation period, {$P_\mathrm{rot} = \SI{122\pm13}{\day}$}, {which} we determined from archival {HATNet{} and ASAS-SN} photometry {combined with newly obtained TJO data}.
}

\maketitle

\section{Introduction}

The Transiting Exoplanet Survey Satellite \citep[TESS;][]{Ricker.2015} has led to the discovery and characterisation of a multitude of small exoplanets. This {growth} was {facilitated by the} intensive spectroscopic follow-up in order to measure radial velocities (RVs) of the TESS objects of interest {\citep[TOI;][]{Guerrero.2021}} and, thus, confirm their planetary nature by measuring their masses. \citep[e.g.][]{Cloutier.2019,Gunther.2019,Luque.2019,AstudilloDefru.2020,Dreizler.2020,Kemmer.2020,Nowak.2020,Stefansson.2020,Soto.2021,Bluhm.2020,Bluhm.2021}. {These discoveries provide valuable data in the ongoing debate as to the origins of super-Earths and mini-Neptunes.}

The so-called radius gap, first observationally {shown} by \cite{Fulton.2017}, divides the transiting sub-Neptunian planets into two different populations. Complementary mass measurements for planets on both sides of the gap confirmed their differing nature: dense and presumably rocky super-Earths with smaller radii and {low-density} enveloped mini-Neptunes with larger radii \citep[e.g.][and references therein]{Rogers.2015,JontofHutter.2019, Bean.2021}. {However, it is not clear which formation mechanisms lead to these distinct planet types.}

{For example, rocky super-Earths could be created by photo-evaporation \citep{Owen.2013,Owen.2017} or core-powered mass loss \citep{Ginzburg.2018,Gupta.2019} of mini-Neptunes with hydrogen-dominated atmospheres. The upper radius and mass limit of the rocky planets and its dependency on the planet's period, or rather {instellation}, are often seen as evidence for this {theory} \citep[e.g.][and references therein]{vanEylen.2018, vanEylen.2021, Bean.2021}. On the other hand, the growing number of planets residing in the gap pose a challenge to this, since their high {instellation} usually excludes substantial H-He atmospheres \citep[e.g.][]{Ment.2019,Dai.2019,Bluhm.2020}. A possible explanation would be the existence of `water planets', which were predicted by classical planet formation models \citep[e.g.][]{Bitsch.2019}. In fact, self-consistent planet formation models, such as those by \cite{Venturini.2020, Venturini.2020b}, predict a bimodal distribution, with purely rocky planets on one side and water-enriched planets on the other. Probing the atmospheres of mini-Neptunes will break the density degeneracy between H-He atmospheres and water planets and provide more insight \citep{Rogers.2010,Lopez.2014, Zeng.2019}. The position of the radius gap, anchored by the distribution of planets on both sides, is also a probe for the underlying principles and an important {input} for theoretical models that aim to describe the formation and evolution of those planets (\citealt{Bean.2021} and references therein).}

{Considered on their own, the planets below the radius gap are also interesting. For example, the increasing statistical sample of the smallest planets will tell us if super-Earths are different from planets with masses and radii similar to Earth, or if the underlying formation mechanisms are the same. Related to this is the question how abundant atmospheres with high mean-molecular-weight similar to the ones that we observe in the Solar System actually are. Particularly suited to answer those questions are the planets orbiting M-dwarf stars, as these will be the first ones where the atmospheric characterisation of exo-Earths will be possible and provide unique insight in {their} structure and composition \citep[e.g.][]{Rauer.2011,Barstow.2016,Morley.2017, Kempton.2018}.}

In this study, we present the discovery of a hot Earth-sized planet orbiting the M3.5\,V-dwarf star, GJ~3929. Based on transit signals observed by TESS, we performed an intensive RV follow-up campaign with CARMENES to confirm its planetary origin. The characterisation of the planet {was} supported by photometric follow-up from SAINT-EX{, LCOGT, and OSN}  that {helped} to refine the transit parameters. Furthermore, we report the detection of a non-transiting{, sub-Neptunian}{-mass} planet candidate with a wider orbit, which is evident in the RV data.

Our paper is organised as follows. In \autoref{sec:data}, we present the{} data used, and in \autoref{sec:stellar_prop} we describe the stellar properties of GJ~3929. Data analysis is described in \autoref{sect:analysis}, and our findings are discussed in \autoref{sec:discussion}. Finally, \autoref{sec:conclusions} gives a summary of our results.

\section{Data}%
\label{sec:data}%

\subsection{High-resolution spectroscopy}
\label{subsec:data_rv}

We took \num{78} observations of GJ~3929 between July 2020 and July 2021 with CARMENES\footnote{Calar Alto high-Resolution search for M dwarfs with {Exoearths} with Near-infrared and visible Echelle Spectrographs, \url{http://carmenes.caha.es}.} \citep{Quirrenbach.2014} in the course of the guaranteed time observation and {legacy} {programmes}. The dual-channel spectrograph covers the spectral ranges of \SIrange[range-phrase={ \text{to} }]{0.52}{0.96}{\micro \meter} in the visible light (VIS; $\mathcal{R} = \num{94600}$) and \SIrange[range-phrase={ \text{to} }]{0.96}{1.71}{\micro \meter} in the near infrared (NIR; $\mathcal{R} = \num{80400}$). For the data reduction and extraction of the RVs, we used \texttt{caracal} \citep{Caballero.2016b} and \texttt{serval} \citep{Zechmeister.2018,Zechmeister.2020}, following our standard approach \citep[e.g.][]{Kemmer.2020}.
{The observations were taken with an exposure time of \SI{30}{\minute}, which resulted in a median signal-to-noise ratio (S/N) of \num{74} ($\text{min.}=15$, $\text{max.}=100$).} After discarding spectra with missing drift correction {(i.e. without simultaneous calibration measurements from the Fabry-P\'erot)}, we retrieved \num{73} RVs in the VIS with a median internal uncertainty of \SI{1.9}{\meter \per \second} and a weighted {root mean square} (wrms) of \SI{4.0}{\meter \per \second} and 72 RVs in the NIR with a median internal uncertainty of \SI{7.3}{\meter \per \second} and wrms \SI{9.1}{\meter \per \second} in the NIR, respectively. Due to the large scatter of the NIR data and the small expected amplitude of the transiting planet candidate, we only considered the VIS data in our analysis \citep{Bauer.2020}.

\subsection{Transit photometry}
\label{subsec:data_transit}

\begin{figure*}
    \centering
    \includegraphics{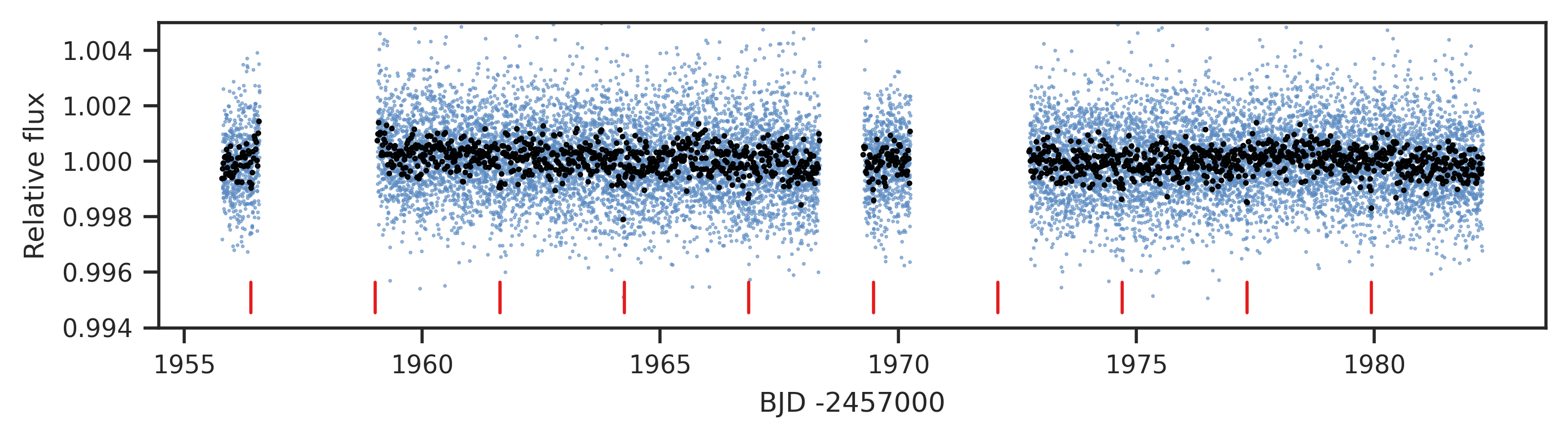}\\
    \includegraphics{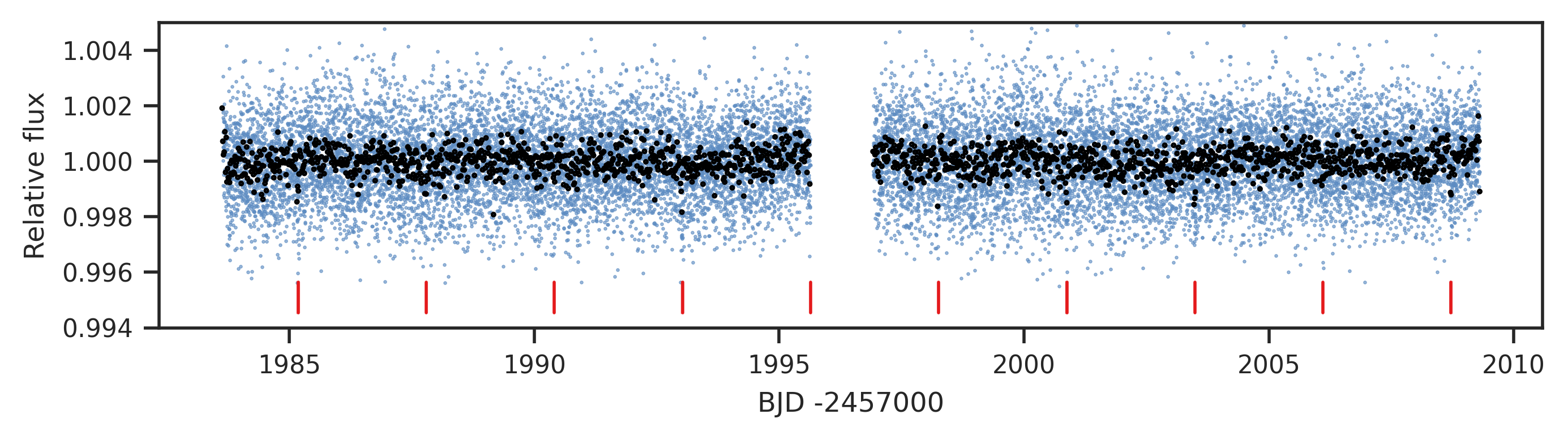}
    \caption{TESS PDCSAP light curves for sector 24 (\emph{top}) and sector 25 (\emph{bottom}). The blue points are the measurements and the black dots are \SI{20}{\minute} bins. The
        transit times of GJ~3929~b are indicated by red ticks.}
    \label{fig:tessfull}
\end{figure*}

For our analysis, we combined the TESS observations with ground based follow-up transit photometry obtained by the {TESS follow-up observing programme sub-group one}. The parameters of the used transit observations are {summarised} in \autoref{tab:transit_phot}. In the following, we provide an overview of the instruments and applied data reduction.

\begin{table*}
    \centering
    \tiny
    \caption{Summary of transit observations.}
    \label{tab:transit_phot}
    \begin{tabular}{l
            c
            c
            S[table-format=3]
            c
            S[table-format=5]
            S[table-format=5]
            S[table-format=2]
            S[table-format=1.2]
            c}
        \hline\hline
        \noalign{\smallskip}
        Telescope  & Date                     & Filter                     & {$t_\text{exp}$} & Airmass\tablefootmark{(a)} & {Duration\tablefootmark{(b)}} & {$N_\text{obs}$} & {Aperture}      & {{10-min rms}} & De-trending                        \\
                   &                          &                            & {[\si{\second}]} &                            & {[min]}                       &                  & {[\si{\pixel}]} & {[ppt]}        &                                    \\
        \noalign{\smallskip}
        \hline
        \noalign{\smallskip}
        TESS S24   & 2020-04-16 to 2020-05-12 & $T$                        & 120              & ...                        & 38138                         & 14650            & 16              & 0.64           & PDC                                \\
        TESS S25   & 2020-05-14 to 2020-06-08 & $T$                        & 120              & ...                        & 36967                         & 17246            & 17              & 0.63           & PDC                                \\
        {OSN}      & 2021-03-14               & None                       & 30               & 1.05$\to$1.00$\to$1.02     & 145                           & 182              & 22              & 0.72           & airmass                            \\
        SAINT-EX   & 2021-03-20               & $I+z$\tablefootmark{{(c)}} & {10}             & 1.64$\to$1.0$\to$1.05      & 334                           & 638              & 11              & 2.19           & PCA                                \\
        LCOGT McD  & 2021-04-10               & $z_{{s}}'$                 & {40}             & 1.54$\to$1.02              & 183                           & 158              & 19              & 0.95           & $totc, width$\tablefootmark{(d)}   \\
        LCOGT CTIO & 2021-04-10               & $z_{{s}}'$                 & {40}             & 2.89$\to$2.40$\to$2.84     & 203                           & 175              & 19              & 1.31           & $totc$, $width$\tablefootmark{(d)} \\
        LCOGT HAL  & 2020-04-15               & $g'$                       & 180              & 1.10$\to$1.03$\to$1.14     & 203                           & 101              & 20              & 1.17           & $totc$, $bjd$\tablefootmark{(d)}   \\
        LCOGT HAL  & 2020-04-15               & $r'$                       & 38               & 1.10$\to$1.03$\to$1.14     & 206                           & 286              & 20              & 0.71           & $totc$, $bjd$\tablefootmark{(d)}   \\
        LCOGT HAL  & 2020-04-15               & $i'$                       & 25               & 1.10$\to$1.03$\to$1.14     & 205                           & 403              & 20              & 0.68           & $sky$, $bjd$\tablefootmark{(d)}    \\
        LCOGT HAL  & 2020-04-15               & $z_{{s}}'$                 & 21               & 1.10$\to$1.03$\to$1.14     & 207                           & 510              & 20              & 0.62           & $width$, $bjd$\tablefootmark{(d)}  \\
        {OSN}      & 2021-05-16               & None                       & 30               & 1.23$\to$1.00              & 180                           & 297              & 22              & 0.55           & airmass                            \\
        {OSN}      & 2021-07-02               & None                       & 30               & 1.03$\to$2.53              & 275                           & 452              & 22              & 1.18           & airmass                            \\
        {OSN}      & 2021-08-13               & None                       & 30               & 1.10$\to$1.60              & 138                           & 229              & 22              & 1.05           & airmass                            \\
        \noalign{\smallskip}
        \hline
    \end{tabular}
    \tablefoot{
        \tablefoottext{a}{The arrows indicate how the airmass has changed over the observation.}
        \tablefoottext{b}{Time span of the observation.}
        \tablefoottext{c}{{Combined range}}
        \tablefoottext{d}{simultaneous to the fits.} Explanation of de-trending parameters: $totc \equiv$ comparison ensemble counts; $width \equiv$ {full width at half maximum (FWHM)} of the target; $bjd \equiv$ BJD timestamp of observation; $sky \equiv$ sky background brightness.}
\end{table*}

\subsubsection{{TESS}}
We retrieved TESS observations for GJ~3929 (TOI\,2013) from the Mikulski Archive for Space Telescopes\footnote{\url{https://mast.stsci.edu}.} for the two sectors 24 and 25 {(see \autoref{fig:tessfull})}. In sector 24 (camera \#1, CCD \#1), one transit event was not observed because of the interruption during the data downlink between $\mathrm{BJD} = 2458968.35$ and $\mathrm{BJD} = 2458969.27$. In sector 25 (Camera \#1, CCD \#2), the measurements were stopped for data download between $\mathrm{BJD} = 2458995.63$ and $\mathrm{BJD} = 2458996.91$, which {led} to one, only partially observed, transit.
We used the pre-search data conditioning simple aperture photometry \citep[PDCSAP;][]{Smith.2012, Stumpe.2012, Stumpe.2014} light curves provided by the Science Processing Operations Center \citep[SPOC;][]{Jenkins.2016}, which are based on simple aperture photometry (SAP) light curves, but further corrected for instrument characteristics. The aperture masks used for retrieving the SAP light curves are shown in \autoref{fig:tpf_plot}.
To reduce the computational cost of the analysis, we {used the extracted transit events only.} In doing so, the baseline for each transit was set to $\pm\SI{3}{\hour}$ with respect to the expected times of transit centre.

\subsubsection{SAINT-EX}
The first follow-up transit photometry for GJ~3929 was taken with the SAINT-EX\footnote{Searching and characterising transiting exoplanets.} {telescope} located at the Observatorio Astron\'omico Nacional in the Sierra de San Pedro M\'artir in Baja California, Mexico. SAINT-EX consists of an Andor iKon-L camera mounted to an \SI{1}{\meter} f/8 Ritchey-Chr\'etien telescope with a pixel scale of {\ang{;;0.34}\,\si{\per\pixel}}, which corresponds to a field of view (FOV) of $\SI{12}{\arcminute}\times \SI{12}{\arcminute}$ \citep{Demory.2020}. The reduction of the data was performed using the instrument's custom pipeline, {\texttt{prince,}} which performs the standard image reduction steps {including} bias, dark, and flat-field correction and provides light curves obtained from differential photometry \citep{Demory.2020}. The light curve used for our analysis was further corrected for systematics using a {principle component ana{ly}sis (PCA)} method based on the light curves of all suitable stars in the FOV except for the target star {\citep{Wells.2021}}.

\subsubsection{LCOGT}
Two transit events were observed with instruments from the LCO\footnote{Las Cumbres Observatory.} global telescope network {\citep[LCOGT;][]{Brown.2013}}. The first one was {observed contemporaneously} by the SINISTRO CCDs at the \SI{1}{\meter} telescopes of the McDonald Observatory (McD) and the Cerro Tololo Interamerican Observatory (CTIO). Both instruments have a pixel scale of {\ang{;;0.389}\,\si{\per\pixel}} and a FOV of $\SI{26}{\arcminute}\times \SI{26}{\arcminute}$. However, the higher airmass at LCOGT CTIO (see \autoref{tab:transit_phot}) led to worse seeing (estimated {point spread function} size {\ang{;;5.34}\,\si{\per\pixel}} vs. {\ang{;;3.97}\,\si{\per\pixel}}) and, therefore, larger scatter of the measurements. The second transit event was observed by the recently commissioned MuSCAT3\footnote{Multicolor Simultaneous Camera for studying Atmospheres of Transiting exoplanets 3.} camera \citep{Narita.2020} mounted on the \SI{2}{\meter} {Faulkes T}elescope North at Haleakal\={a} Observatory (HAL). MuSCAT3 operates simultaneously in the four passbands, $g'$, $r'$, $i'$, and $z_{{s}}',$ and has a pixel scale of {\ang{;;0.27}\,\si{\per\pixel}} corresponding to a FOV of ${\ang{;9.1;}}\times {\ang{;9.1;}}$. {All LCOGT observations were calibrated by the standard LCOGT {\tt BANZAI} pipeline \citep{McCully.2018, McCully.2018b}, and photometric data were extracted using {\tt AstroImageJ} \citep{Collins.2017}.}

\subsubsection{OSN}
We detected four transit events with the \SI{90}{\centi\meter} Ritchey-Chr\'etien telescope of the Observatorio de Sierra Nevada (OSN). The telescope is equipped with a $\text{2k}\times\text{2k}$ CCD camera with a pixel scale of \ang{;;0.4}\,\si{\per\pixel} that provides a FOV of $\ang{;13.2;} \times \ang{;13.2;}$ \citep{Amado.2021}. All frames were corrected for bias and flat-fielding, and the light curves were obtained using synthetic aperture photometry. Outliers due to bad weather or very high airmass were removed from the dataset. The light curves were de-trended before the fitting using the airmass in a linear fit.

\begin{figure}
    \center
    \includegraphics[width=\columnwidth]{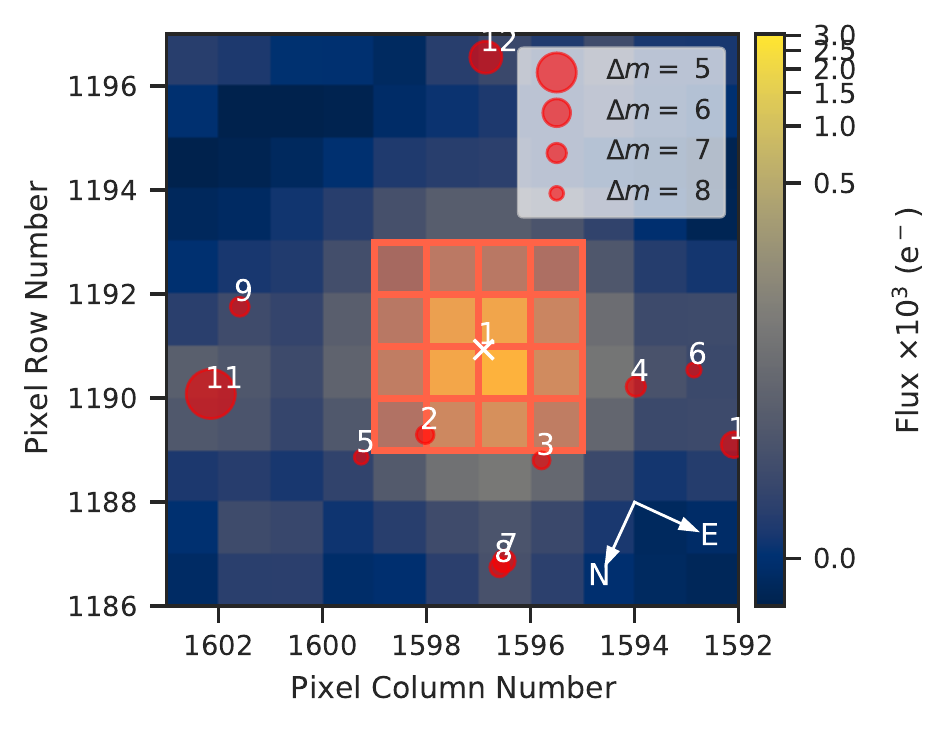}
    \includegraphics[width=\columnwidth]{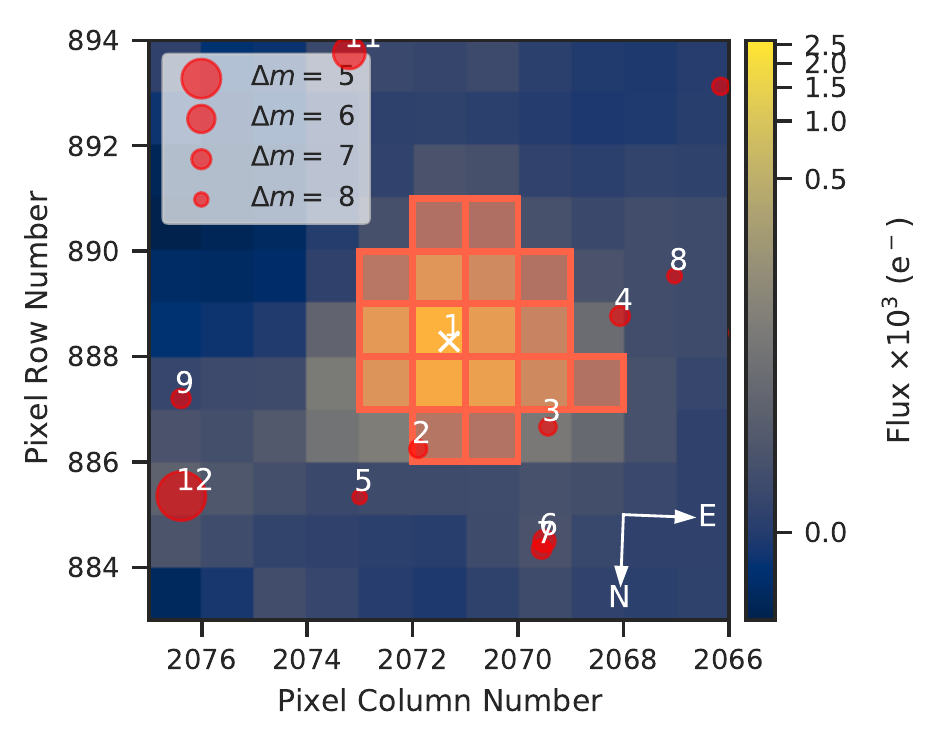}
    \caption{TESS TPFs of GJ~3929. \textit{Top:} TESS sector 24, \textit{bottom:} TESS sector 25.
        The position of GJ~3929 is denoted by a white cross, and the aperture mask used to create the PDCSAP light curves is shown as the pixels with orange borders. For comparison, nearby sources from the \emph{Gaia} DR2 catalogue \citep{GaiaCollaboration.2018}, up to a difference of $\Delta m = \SI{8}{mag}$ in brightness compared to GJ~3929, are plotted by red circles. Figure created using \texttt{tpfplotter} \citep{Aller.2020}.}
    \label{fig:tpf_plot}
\end{figure}

\subsection{Long-term photometry}

In addition to the photometric transit observations, we used long-term photometry to determine the stellar rotation period.

\subsubsection{HATNet}
The photometric variability of GJ~3929 was previously investigated by \cite{Hartman.2011} using data from  the HATNet telescope network \citep{Bakos.2004,Bakos.2006}. HATNet comprises a network of six cameras attached to \SI{11}{\centi\meter} telescopes located in Arizona and Hawai'i. The cameras have a FOV of $\ang{8.2}\times\ang{8.2}$ and a pixel scale of \SI{14}{\arcsecond\per\pixel}. We retrieved the observations covering a time span of \SI{200}{\day} between December 2004 and July 2005 from the NASA Exoplanet Archive\footnote{\url{https://exoplanetarchive.ipac.caltech.edu/docs/datasethelp/ETSS_HATNet.html}.}. The data were taken by telescopes \#9 and \#11 in the Cousins $I_c$ filter at the Fred Lawrence Whipple Observatory in Arizona. Originally, the data were taken with a cadence of \SI{5.5}{\minute}, but for our search for long-periodic signals we used the nightly binned values. In this way, we obtain a mean uncertainty of \SI{1.24}{ppt} and rms of \SI{2.19}{ppt}.

\subsubsection{ASAS-SN}
We obtained more than \SI{5}{\year} of archival data from ASAS-SN \citep{Shappee.2014,Kochanek.2017}, {which} were taken between April 2013 and September 2018. ASAS-SN currently consists of 24 cameras mounted on the \SI{14}{\centi\meter} Nikon telephoto lenses at six different sites around the globe. Each unit has a FOV of $\ang{4.5}\times\ang{4.5}$ with a pixel scale of \ang{;;8.0}\,\si{\per\pixel}. The observations of GJ~3929 were obtained in the $V$ band with the second camera in Hawai'i and have a mean uncertainty of \SI{4.82}{ppt} and rms of \SI{6.68}{ppt}.

\subsubsection{TJO}
We observed GJ~3929 from April to October 2021 with the \SI{0.8}{\meter} Joan Oró telescope \citep[TJO][]{Colome.2010} at the Montsec Observatory in Lleida, Spain. We obtained a total of 593 images with an exposure time of \SI{60}{\second} using the Johnson $R$ filter of the LAIA imager, a $\text{4k}\times\text{4k}$ CCD with a field of view of \ang{;30;} and a scale of \ang{;;0.4}\,\si{\per\pixel}. The images were calibrated with darks, bias, and flat fields with the {\texttt{icat}} pipeline of the TJO \citep{Colome.2006}. The differential photometry was extracted with {\texttt{AstroImageJ}} using the aperture size that minimised the rms of the resulting relative fluxes, and a selection of the ten brightest comparison stars in the field that did not show variability. Then, we used our own pipelines to remove outliers and measurements affected by poor observing conditions or presenting a low signal-to-noise ratio. For our analysis, we binned the data nightly, which resulted in a total of \num{54} measurements with a mean uncertainty of \SI{1.90}{ppt} and rms of \SI{6.22}{ppt}.

\subsection{High-resolution imaging}
\label{subsec:hri}
We observed GJ~3929 with the high-spatial resolution camera AstraLux \citep{Hormuth.2008}, which is located at the \SI{2.2}{\meter} telescope of the Calar Alto Observatory (Almer\'ia, Spain). The observations were carried out on 7 August 2020 at an airmass of 1.1 and under moderate weather conditions with a mean seeing of {\ang{;;1.1}}.
In total, we obtained \num{93700} frames in the Sloan Digital Sky Survey $z'$ filter (SDSS$z$) with \SI{10}{\milli\second} exposure times and windowed to a FOV of \SI{6 x 6}{\arcsecond}. We used the instrument pipeline to select the \SI{10}{\percent} frames with the highest Strehl ratio \citep{Strehl.1902} and to combine them into a final high-spatial-resolution image. Based on this final image, a sensitivity curve was computed using our own developed \texttt{astrasens}\footnote{\url{https://github.com/jlillo/astrasens}.} package \citep{LilloBox.2012, LilloBox.2014}.

\section{Properties of {GJ~3929}}
\label{sec:stellar_prop}

\begin{table}[!ht]
    \caption{Stellar parameters of {GJ~3929}.}
    \label{tab:stellar_parameters}
    \centering
    \begin{tabular}{l c r}
        \hline\hline
        Parameter                                     & Value                 & Ref.               \\
        \hline
        \noalign{\smallskip}
        \multicolumn{3}{c}{\em Name and identifiers}                                               \\
        \noalign{\smallskip}
        {Name}                                        & {GJ~3929}             & {Gli91}            \\
        {Alternative name}                            & {G~180--18}           & {Gic59}            \\
        Karmn                                         & J15583+354            & Cab16              \\
        TIC                                           & 188589164             & Stas19             \\
        TOI                                           & 2013                  & Gue21              \\
        {\emph{Gaia} EDR3}                            & {1372215976327300480} & {\emph{Gaia} EDR3} \\
        \noalign{\smallskip}
        \multicolumn{3}{c}{\em Coordinates, magnitudes, and spectral type}                         \\
        \noalign{\smallskip}
        $\alpha$ {(epoch 2016.0)}                     & 15\,58\,18.80         & \emph{Gaia} EDR3   \\
        $\delta$ {(epoch 2016.0)}                     & +35\,24\,24.3         & \emph{Gaia} EDR3   \\
        Spectral type                                 & M3.5\,V               & L{\'e}p13          \\
        $T$\,[mag]\tablefootmark{(a)}                 & $10.2705\pm0.0074$    & {Stas19}           \\
        \noalign{\smallskip}
        \multicolumn{3}{c}{\em Parallax and kinematics}                                            \\
        \noalign{\smallskip}
        $\mu_{\alpha}\cos\delta$\,[\si{mas\,yr^{-1}}] & $-143.06\pm0.02$      & \emph{Gaia} EDR3   \\
        $\mu_\delta$\,[\si{mas\,yr^{-1}}]             & $318.12\pm0.03$       & \emph{Gaia} EDR3   \\
        $\pi$\,[\si{mas}]                             & $63.173\pm0.020$      & \emph{Gaia} EDR3   \\
        $d$\,[\si{pc}]                                & $15.830\pm0.006$      & \emph{Gaia} EDR3   \\
        $\gamma$ [\si{\kilo\meter\per\second}]        & $+9.54\pm0.01$        & Jeff18             \\
        $U$\,[\si{\kilo\meter\per\second}]            & $-21.05\pm0.04$       & {This work}        \\
        $V$\,[\si{\kilo\meter\per\second}]            & $+10.85\pm0.06$       & {This work}        \\
        $W$\,[\si{\kilo\meter\per\second}]            & $+14.66\pm0.08$       & {This work}        \\
        RUWE                                          & 1.19                  & \emph{Gaia} EDR3   \\
        \noalign{\smallskip}
        \multicolumn{3}{c}{\em Photospheric parameters}                                            \\
        \noalign{\smallskip}
        $T_\tx{eff}$\,[K]                             & $3369\pm51$           & This work          \\
        $\log{g}$\,[dex]                              & $4.84\pm0.04$         & This work          \\
        $[\tx{Fe/H}]$\,[dex]                          & $+0.00\pm0.16$        & This work          \\
        \noalign{\smallskip}
        \multicolumn{3}{c}{\em Physical parameters}                                                \\
        \noalign{\smallskip}
        $L_\star$\,[{$L_\odot$}]                      & {$0.01155\pm0.00011$} & {Cif20}            \\
        $R_\star$\,[$R_\odot$]                        & $0.315\pm0.010$       & This work          \\
        $M_\star$\,[$M_\odot$]                        & $0.309\pm0.014$       & This work          \\
        \noalign{\smallskip}
        \multicolumn{3}{c}{\em Activity parameters}                                                \\
        pEW (H$\alpha$) [\si{\angstrom}]              & $-0.029\pm0.031$      & This work          \\
        $v\sin{i}$\,[\si{\kilo\meter\per\second}]     & $< 2$                 & This work          \\
        $P_\mathrm{rot}$ [d]                          & {$122\pm13$}          & This work          \\
        \noalign{\smallskip}
        \hline
    \end{tabular}
    \tablefoot{\tablefoottext{a}{{Additional photometric passbands are listed in \autoref{tab:photometric_passbands}.}}}
    \tablebib{
    {Gli91: \citet{Gliese.1991};}
        {Gic59: \citet{Giclas.1959};}
    Cab16: \citet{Caballero.2016};
    Stas19: \citet{Stassun.2019};
    {Gue21: \citet{Guerrero.2021};}
    \emph{Gaia} EDR3: \citet{GaiaCollaboration.2021};
    L{\'e}p13: \citet{Lepine.2013};
    2MASS: \citet{Skrutskie.2006};
    Jeff18: \citet{Jeffers.2018};
    Cif20: \citet{Cifuentes.2020}.
    }
\end{table}

The star GJ~3929 (G~180-18, Karmn J15583+354) is located at a distance of only \SI{15.830\pm0.006}{\parsec} and shows a high proper motion \citep{Schneider.2016, GaiaCollaboration.2018}. \cite{Lepine.2013} classified the star as {an} M3.5\,V red dwarf. We calculated homogeneous stellar parameters from the CARMENES high-resolution spectra using our standard method: {the} luminosity, $L_\star= \SI{0.01155\pm0.00011}{L_\odot}$, {were} determined in \cite{Cifuentes.2020}. Following \cite{Passegger.2019}, and assuming $v \sin i = \SI{2}{\kilo\meter \per\second}$, we derived the effective temperature, $T_\mathrm{eff}=\SI{3369\pm51}{\kelvin}$, surface gravity $\log g = \SI{4.84\pm0.04}{dex}$, and metallicity $[\mathrm{Fe/H}] = \SI{0.00\pm0.16}{dex}$ using the VIS spectra\footnote{{These astrophysical stellar parameters agree within $1 \sigma$ uncertainties with those recently published by \citet{Marfil.2021}.}}. Finally, we computed the stellar radius, $R_\star=\SI{0.315\pm0.010}{R_\odot,}$ using the Stefan-Boltzman law, and, consequentially, the mass, $M_\star=\SI{0.309\pm0.014}{R_\odot,}$ from the empirical mass-radius relation for M dwarfs of \citet{Schweitzer.2019}. {Additionally, we computed galactocentric space velocities $UVW$ as in \citet{CortesContreras.2017}.}

{From the analysis of the H$\alpha$ pseudo-equivalent width (pEW), we found that GJ~3929 is an H$\alpha$-inactive star and is consistent with the previous results of \cite{Schofer.2019} and \cite{Jeffers.2018}. In addition, we investigated if there are any correlations between the measured CARMENES RV values and all of the activity indices using the Pearson’s $r$ coefficient where a value of $>0.7$ or $<-0.7$ indicates strong correlation or anti-correlation as previously described by \cite{Jeffers.2020}. We found no strong or even moderate correlations between the measured CARMENES RVs and the {activity} indices, confirming that GJ~3929 is a magnetically inactive star.} In \autoref{subsec:activity}, we present a combined analysis of CARMENES activity indicators and photometry from HATNet{, ASAS-SN, and TJO}, from which we determine a stellar rotation period of {\SI{122\pm13}{\day}}.

GJ~3929 has no known stellar companions. {From} the high-resolution imaging presented in \autoref{subsec:contamination}, we can rule out companions up to contrasts of $\Delta m=\SI{4}{mag}$ down to a separation of {\ang{;;0.2}} and $\Delta m=\SI{5.5}{mag}$ for separations of {\ang{;;0.4}} to {\ang{;;2}}. {Additionally, {\em Gaia} provides a re-normalised unit weight error of 1.19, which is below the critical value of 1.41 that would hint to a close companion. Besides this, we complemented the multiplicity analysis with a wide common-proper-motion companion search with {\em Gaia} EDR3 data up to a projected physical separation of $10^5$\,au (over 10\,arcmin in angular separation); no wide companions with similar parallax and proper motion were found.} Furthermore, the astrometric excess-noise is \SI{0.22}{\milli as}, which is consistent with the jitter of other sources with comparable $G$ magnitudes between \SIrange{11}{13}{mag}. Our RV analysis in \autoref{subsec:rv_only} also shows no signals that would indicate any massive companions. A summary of the compiled stellar parameters and their sources is provided in \autoref{tab:stellar_parameters}.

\section{Analysis and results}
\label{sect:analysis}

\subsection{{Transit detection by the SPOC}}
{The SPOC investigated the PDCSAP flux time series for sector 24 with the Transiting Planet Search \citep{Jenkins.2002b,Jenkins.2010,Jenkins.2020} module using an adaptive, wavelet-based matched filter, which detected transit events with a period of \SI{\sim2.6}{\day} and generated a
    `threshold crossing event'. The data were fitted with an initial limb-darkened transit model \citep{Li.2019} and subjected to a suite of diagnostic tests to help elucidate the nature of the signal \citep{Twicken.2018}. The transit signal passed all the tests in the {data validation} module, and it was promoted from `threshold crossing event' to TOI status by the TESS Science Office on 19 June 2020 after reviewing the Data Validation reports \citep{Guerrero.2021}. Subsequent joint analyses of sectors 24 and 25 indicated that the transit source is located within {$\ang{;;2.8}\pm \ang{;;6.6}$} of GJ 3929. The multiple transiting planet search failed to identify any additional transiting planet signatures.}

\subsection{Limits of photometric contamination}
\label{subsec:contamination}

\begin{figure}
    \centering
    \includegraphics[width=\columnwidth]{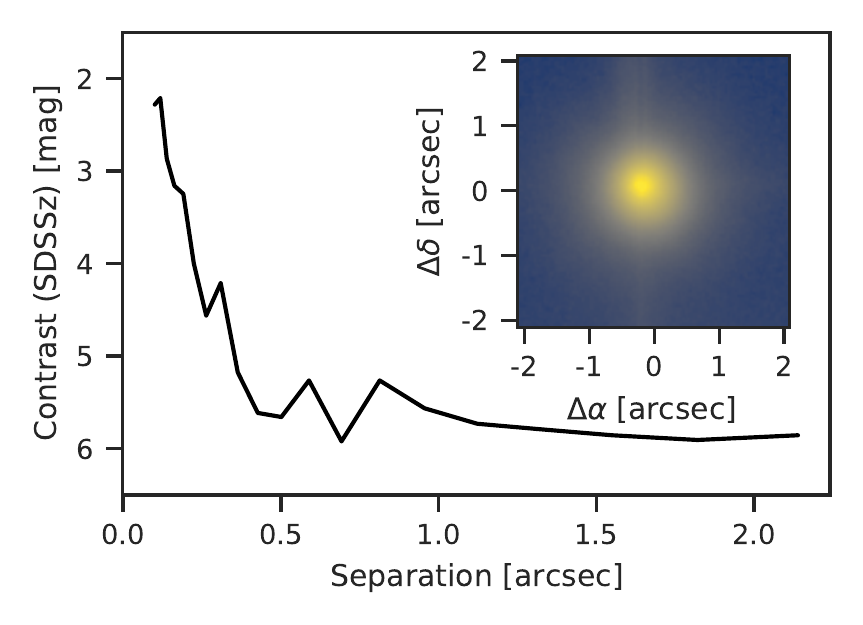}
    \caption{Contrast curve of the AstraLux high-resolution image. The image used to create the contrast curve is shown in the inset.}
    \label{fig:astralux}
\end{figure}

{As seen in \autoref{fig:tpf_plot}}, there are no \emph{Gaia} sources {down} to a brightness difference of $\Delta G \approx \SI{7}{mag}$  within the apertures used for creating the SAP light curves. {The SPOC estimated a contamination in the photometric aperture of about \SI{0.4}{\percent} for both sectors, based on the crowding and the location of the target star on the CCD using the pixel response functions reconstructed from data collected during commissioning and early science operations.} Nevertheless, we obtained additional lucky imaging observations to rule out contamination of the light curves by bound or unbound companions at sub-arcsecond separations {(\autoref{subsec:hri})}. The AstraLux image of GJ~3929 and the contrast curve created from it {are shown} in \autoref{fig:astralux}. We find no evidence of additional sources within this FOV and within the computed sensitivity limit. This allows us to set an upper limit to the contamination in the light curve of around \SI{0.4}{\percent} down to {\ang{;;0.4}} and \SI{2.5}{\percent} down to {\ang{;;0.2}}.

{Analogously} to \cite{LilloBox.2014}, we further used the contrast curve to estimate the probability of contamination from blended sources in the TESS aperture based on the TRILEGAL\footnote{\url{http://stev.oapd.inaf.it/cgi-bin/trilegal}.} Galactic model \citep[v1.6{,}][]{Girardi.2012}. The transiting planet candidate around GJ~3929 produces a signal that could be mimicked by blended eclipsing binaries with magnitude contrasts up to  $\Delta m_{\rm b,max} \approx \SI{7.3}{mag}$ in the SDSS$z$ passband. {Translating this contrast} results in a low probability of \SI{0.1}{\percent} for an undetected source, and an even lower probability of such a source being an appropriate eclipsing binary.
Given these numbers, we assumed that the transit signal is not due to a blended binary star and that the probability of a contaminating source is nearly zero.

\subsection{{Modelling technique}}
We used \texttt{juliet}\footnote{\url{https://juliet.readthedocs.io/en/latest/}.} \citep{Espinoza.2019} for the analysis and modelling of the transit and RV data. Thereby, we follow our method as detailed for example by \cite{Luque.2019}, \cite{Kemmer.2020}, or  \cite{Stock.2020}. Because of the variety of {instruments used} and the large dataset, it would not be reasonable to perform the model selection on the combined RV and transit data. Therefore, in the following we first present {transit-only and RV-only} analyses to determine the individually best fitting models, which were later combined into a joint fit to retrieve the most precise parameters for the system.

\subsection{Transit-only modelling}
\label{subsec:transit_only}

In the first step of the modelling, we combined the TESS light curves with the SAINT-EX{, LCOGT, and OSN} follow-up transits to obtain a very precise updated ephemeris of the transiting planet candidate, {which} was later used as prior information for the RV-only modelling.

\paragraph{Planet parameters.} Based on the analysis of the TESS light curves by the SPOC pipeline \citep{Li.2019}, the transiting planet candidate has a period of \SI{2.616277\pm0.000113}{\day}. We used this information to set a uniform prior between \SIrange{2}{3}{\day} for our analysis. The {time-of-transit} centre was chosen accordingly to be uniform between BJD \SIrange{2459319.0}{2459322.0}{\day}, {which comprises the decently resolved follow-up transit observed by LCO{GT}-HAL.} Following our usual approach \citep[e.g.][]{Luque.2019,Kemmer.2020,Bluhm.2021}, we fitted for the stellar density, $\rho_*$, instead of the scaled planetary semi-major axis, $a/R_*$. In doing so, we used a {normally} distributed prior centred on the density calculated from the parameters in \autoref{tab:stellar_parameters}. For this, we assigned a width of three times the propagated uncertainty. Furthermore, we implemented the re-parameterised fit variables $r_1$ and $r_2$, which replace the planet-to-star radius ratio, $p$, and the impact parameter, $b$, and allow for a uniform sampling between zero and one \citep{Espinoza.2018}. Since the information content regarding the eccentricity is rather small for the light curves \citep{Barnes.2007, Kipping.2008, VanEylen.2015}, we assumed it to be zero for the transit-only modelling. Constraints on the eccentricity were later investigated using the RV data (see \autoref{subsec:rv_only}).

\paragraph{Instrument parameters.}
The analysis of the high-resolution images (\autoref{subsec:contamination}) did not indicate any contaminating sources within the apertures that were used to generate the light curves. Therefore, the dilution factor was fixed to 1 for all instruments. {Following \cite{Espinoza.2015}, we used a quadratic limb-darkening law for the space-based TESS light curves, parameterised by $q_1$ and $q_2$ as in \cite{Kipping.2013}. The parameters were shared between the two sectors.} For all the other ground-based follow-up observations, we assumed a linear limb darkening {with coefficient} $q$. The offsets between the instruments, $mflux$, were assumed to be {normally} distributed around 0 with a standard deviation of 0.1, whereas the additional scatter that was added in quadrature to the nominal uncertainty values was {log-uniformly} distributed between \SIrange{0}{5000}{ppm}. The light curves from the LCOGT were de-trended simultaneously with the fits{, whereas, following a preliminary analysis, de-trending of the SAINT-EX light curve did not bring any improvement, which is why we refrained from doing so in the analysis}. {Moreover, the OSN light curves were de-trended before the fit as described in \autoref{subsec:data_transit}}.
We invite the reader to consult \autoref{tab:transit_phot} for an overview of the used de-trending parameters of the individual light curves. {In this way}, we determined a refined period of  {$P=\SI[separate-uncertainty=false]{2.6162733\pm0.0000034}{\day}$} and {$t_0=\SI[separate-uncertainty=false]{2459320.05803\pm0.00024}{\day}$} from the fit.

In order to search for additional transit signals in the data, we applied the model from this fit to the entire TESS dataset (i.e. uncropped) and ran a transit least squares \citep[TLS;][]{Hippke.2019} periodogram on the residuals. The periodogram did not show any further significant signals.

\subsection{RV-only modelling}
\label{subsec:rv_only}

\begin{figure}
    \centering
    \includegraphics{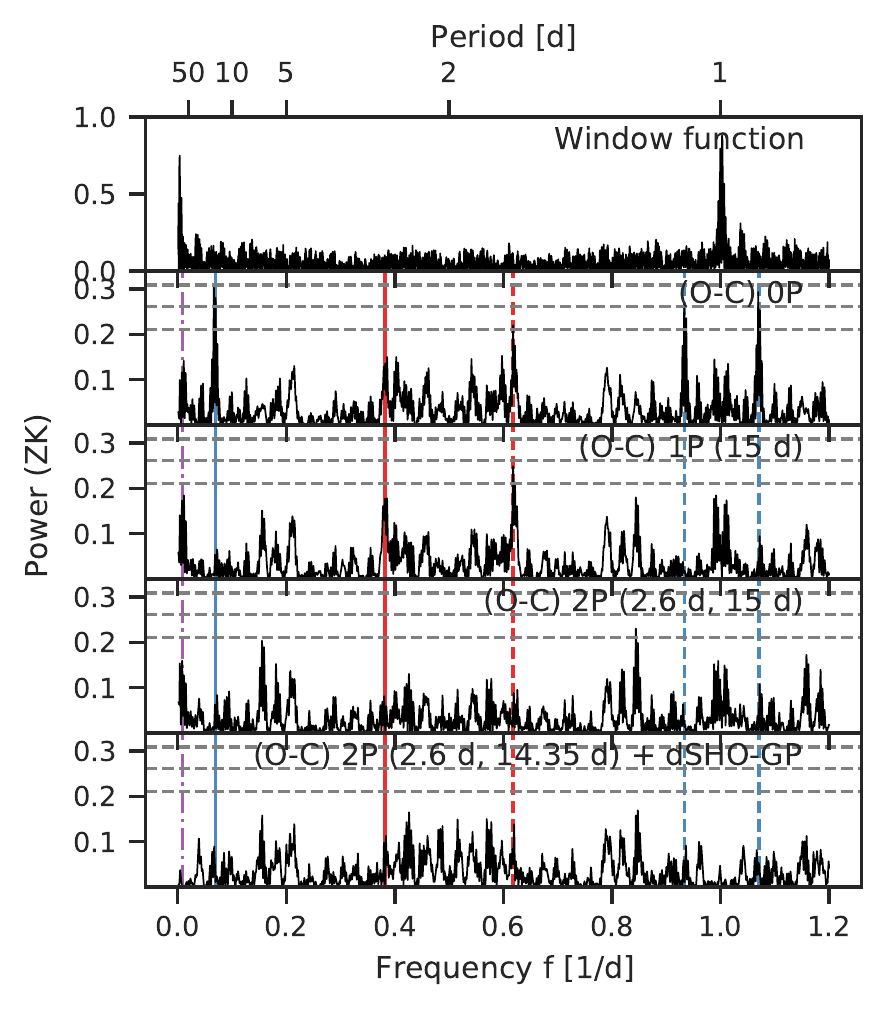}
    \caption{GLS periodogram analysis of the RVs. {In the first panel, we show the window function of the CARMENES data. In the subsequent panels, the residuals after subtracting models of increasing complexity are presented. The components that were considered for the fits are listed in the inset texts (see also \autoref{tab:rv_evidence}).} The period, $P=\SI{2.62}{\day}$, and one-day alias, $P=\SI{1.62}{\day}$, of the transiting planet are marked by the red solid and dashed lines, while the $\sim15$-day periodicity and its {daily} aliases are marked by blue solid and dashed lines, respectively. Additionally, even though insignificant in the periodogram, the stellar rotation period {of} $P=\SI{122}{\day}$ (\autoref{subsec:activity}) is indicated by the purple dot-dashed line. We {normalised} the power using the {parametrisation} of \cite{Zechmeister.2009}, and the 10, 1, and \SI{0.1}{\percent} FAPs denoted by the horizontal grey dashed lines were calculated using the analytic expression.}
    \label{fig:gls_rv}
\end{figure}

\subsubsection{Periodogram analysis}
We used {generalised} Lomb-Scargle periodograms \citep[GLS;][]{Zechmeister.2009} implemented in \texttt{Exo-Striker} \citep{Trifonov.2019, Trifonov.2021} to identify prominent signals in the RV data, as illustrated by \autoref{fig:gls_rv}. The dominant period is not that of the transiting planet candidate at about \SI{2.6}{\day}, but a signal with periodicity of $P\approx\SI{15}{\day}$ and its one-day aliases. Furthermore, aliasing due to the seasonal observability of GJ~3929 ($f_s\approx\SI{1/292}{\per\day}$) {splits} the $\sim$15-day signal up into multiple close peaks {by} itself. The two prominent peaks are thereby at periods of $P\approx\SI{14.3}{\day}$ and $P\approx\SI{15.0}{\day}$ (see also \autoref{subsubsec:aliasing}). As there are no obvious indications of a transiting signal corresponding to these two periodicities that would help to distinguish the aliases, we {used a sinusoidal fit with an uninformative period boundary between \SIrange{10}{20}{\day} to subtract the signal, and determined a period of $P=\SI{15.03}{\day}$.} The residuals of this fit show a peak {with} about a \SI{2}{\percent} false alarm probability (FAP) at a period of $P\approx\SI{1.62}{\day}$. This period corresponds to the one-day alias of the \twoday signal seen in the transits, which is {itself} apparent {only} as {an insignificant} signal in the GLS periodogram. The photometric observations presented in the previous section, however, supplied precise information on the period and transit time, and hence phase, of the transiting planet candidate. We therefore simultaneously fitted the $\sim$15-day signal in combination with a sinusoid of $P\approx\SI{2.62}{\day,}$ whose ephemeris was fixed to the values from \autoref{subsec:transit_only}. The residuals of this fit do not show any power at the period of $P\approx\SI{1.62}{\day}$, which confirms that the peak is indeed correlated in phase with the signal of the transiting planet candidate, and, thus, it is caused by aliasing. {Even though never significant, a peak near the stellar rotation period of \SI{122}{\day} is {also} visible in the GLS periodograms (see \autoref{subsec:activity}).}

\begin{figure}
    \centering
    \includegraphics{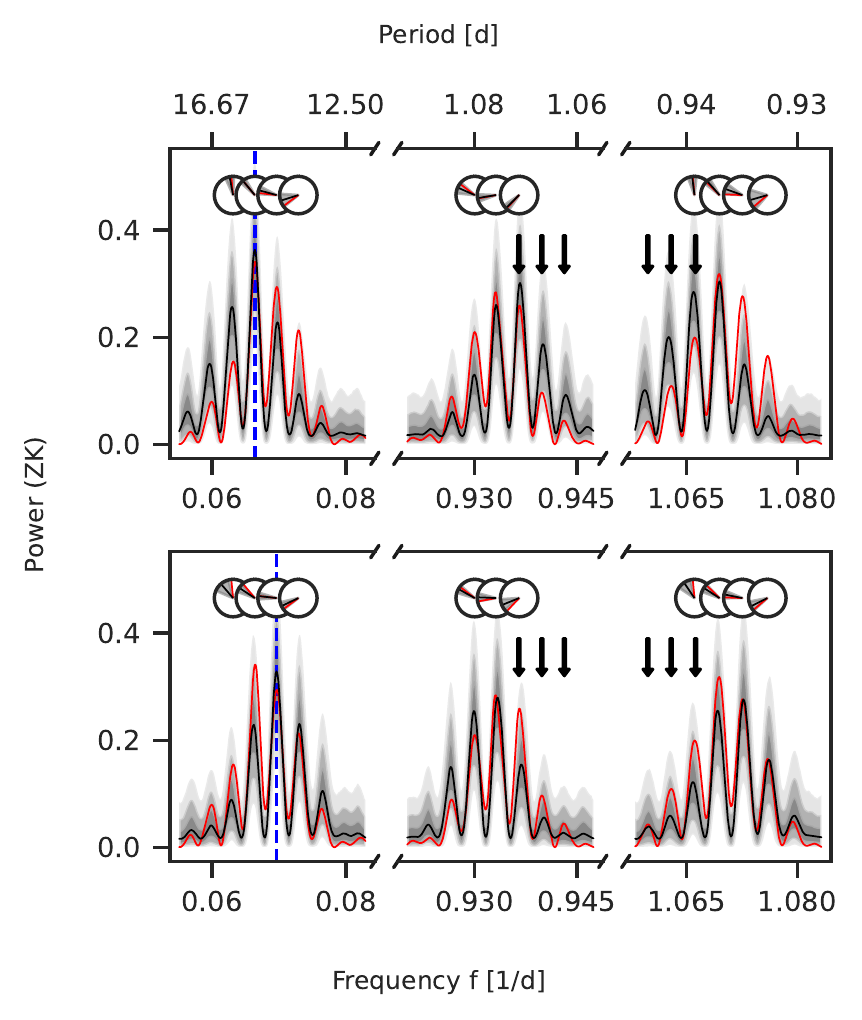}
    \caption{Alias test for {the} \fourteenday and \fifteenday periods using \texttt{AliasFinder}. We generated \num{5000} synthetic datasets for each period to produce synthetic periodograms (black lines){,} which are compared with the periodogram of the observed data (red lines). The simulation for the \fifteenday signal is shown in the top row and the simulation for the \fourteenday signal in the bottom row, each period indicated by a vertical blue dashed line, respectively. Black lines depict the median of the samples for each simulation, and the grey shaded areas are the 50, 90, and \SI{99}{\percent} confidence intervals. Furthermore, the phases of the peaks as {determined by} the GLS {periodogram} are displayed in the circles, following the same colour scheme (the grey shades denote the standard deviations of the simulated peaks). {The black arrows point out the difference in the periodograms for the {daily} aliases that allows to {identify} the best matching period (see the text for the discussion).}}
    \label{fig:aliasfinder}
\end{figure}

\subsubsection{Determining the true period underlying the \SI{\sim15}{\day} {GLS peaks}}
\label{subsubsec:aliasing}
{We made use of the \texttt{AliasFinder}\footnote{\url{https://github.com/JonasKemmer/AliasFinder}.}  \citep{Stock.2020c, Stock.2020b} to} identify the true period underlying the {GLS peaks} of  \SI{14.3}{\day} and \SI{15.0}{\day}{, which are aliases of each other} caused by the $f_s\approx \SI{1/292}{\per\day}$ sampling frequency. The script implements the principle of \cite{Dawson.2010} and allows us to visually compare the observed periodogram with synthetic periodograms originating from different possible alias frequencies. In doing so, we excluded the influence of the \twoday signal by first removing it from the data with a sine fit, as we did for the periodogram analysis.
\autoref{fig:aliasfinder} shows the resulting comparison periodograms for the 14.3-day and 15.0-day periods. Each panel shows three sections of the full periodogram; the first panel is the region around \SI{\sim15}{\day,} which highlights the aliasing due to the \SI{\sim292}{\day} sampling ($f_\text{alias}=|f\pm \frac{1}{\SI{292}{\day}}|$), and the other two show the aliases of the daily sampling (\textit{middle panel}: $f_\text{alias}=|f-\frac{1}{\SI{1}{\day}}|)$, \textit{right panel}: $f_\text{alias}=|f+\frac{1}{\SI{1}{\day}}|)$). The idea behind this is that the true frequency should be able to explain both patterns well, as they are generated independently by it.

While the phases originating from the stimulation of the \fifteenday period show less deviations from the observed phases than {those} from the \fourteenday period (see the circles in \autoref{fig:aliasfinder}), the evaluation of the periodograms {implies} that \SI{14.3}{\day} is the period underlying our data. The peaks originating from the \fifteenday period show a shifted distribution when compared to the observed periodogram, which can be seen especially well in the {daily} aliases. There, the {envelope of the} aliases at {\SI{\sim1.07}{\day}} {is} shifted towards shorter periods, and {those} at \SI{\sim0.94}{\day} are shifted towards longer periods -as is expected when the simulated period is larger than the underlying one {(see the black arrows in \autoref{fig:aliasfinder})}. The distribution of the simulated periodogram originating from the \fourteenday period, on the other {hand}, follows the observed periodogram. This is also reflected by the rms power of the residuals after subtracting the median GLS model from the observed one, where we found a value of \num{7.12} for the \fourteenday period and \num{7.97} for the \fifteenday period. We also got the same result if we generated the periodograms from the posterior samples of the fits, as shown in \autoref{app:alt_alias}.
Therefore, we concluded that the \fourteenday period is the true period of the {$\sim15$-day} signal, and adopted from then on a uniform prior corresponding to the peak width in the periodogram between \SIrange{13.98}{14.71}{\day} whenever this signal was considered in a fit.
Using the \fourteenday period for pre-whitening the periodogram instead of the uninformative prior improved the FAP of the transiting planet candidate's alias in the residual GLS periodogram to \SI{0.8}{\percent}.

\subsubsection{Significance of the transiting planet candidate in the RVs}
\label{subsec:significance_period}

{In the next step, we derived the FAP for the signal in the RVs to occur exactly at the period of the transiting planet candidate.} {One problem is the strong aliasing, which has the consequence that the \oneday alias of the transit signal has in fact the highest power in the periodogram. For our approach, we used the randomisation method discussed by \cite{Hatzes.2019}, \cite{Luque.2019}, \cite{Kemmer.2020}, or \cite{Bluhm.2021}, where the FAP is determined for increasingly smaller frequency ranges around the period in question and extrapolated with a third-order polynomial fit to a window size of zero. To account for the aliasing in our case, we considered two windows that comprise the two peaks {at the periods of \SI{2.62}{\day} and \SI{1.62}{\day}} in the periodogram and compared their combined power with the combined power of the respective highest peaks within the two windows.} In doing so, we found a FAP of \SI{0.1}{\percent} (\autoref{fig:fap}). We therefore concluded that we detected a genuine signal of the transiting planet candidate in the RV measurements.

\subsubsection{Model comparison}
\begin{table}
    \caption{Model comparison for RVs based on Bayesian log evidence.}
    \label{tab:rv_evidence}
    \centering
    \begin{tabular}{l S[table-format=+3.1] S}
        \hline\hline
        Model                                                             & $\ln\mathcal{Z}$  & $\Delta\ln\mathcal{Z}$ \\
        \hline
        \noalign{\smallskip}
        \multicolumn{3}{l}{\em No planet:}                                                                             \\
        \noalign{\smallskip}
        0P                                                                & -213.3            & -6.0                   \\
        \noalign{\smallskip}
        \multicolumn{3}{l}{\em {Two-signal models (without activity modelling)}:}                                      \\
        \noalign{\smallskip}
        2P$_\text{({2.6} d, 14.3 d)}$                                     & -211.2            & -3.9                   \\
        2P$_\text{({2.6} d-ecc, 14.3 d)}$                                 & -211.8            & -4.5                   \\
        2P$_\text{{2.6} d, 14.3 d-ecc)}$                                  & -211.2            & -3.9                   \\
        2P$_\text{({2.6} d-ecc, 14.3 d-ecc)}$                             & -212.1            & -4.8                   \\
        \noalign{\smallskip}
        \multicolumn{3}{l}{\em {Three-signal models (with activity modelling)}:}                                       \\
        \noalign{\smallskip}
        \textbf{2P$_\text{({2.6} d, 14.3 d)}$ + dSHO-GP$_\text{(120 d)}$} & $\mathbf{-207.3}$ & $\mathbf{0.0}$         \\
        2P$_\text{({2.6} d-ecc, 14.3 d)}$ + dSHO-GP$_\text{(120 d)}$      & -207.7            & -0.4                   \\
        2P$_\text{({2.6} d, 14.3 d-ecc)}$ + dSHO-GP$_\text{(120 d)}$      & -208.0            & -0.7                   \\
        2P$_\text{({2.6} d-ecc, 14.3 d-ecc)}$ + dSHO-GP$_\text{(120 d)}$  & -208.7            & -1.4                   \\
        \hline
    \end{tabular}
    \tablefoot{{The bold font denotes the model that was used in the joint fit.}}
\end{table}

\paragraph{Planet and instrument parameters.}
The periodogram and alias analysis showed two relevant periodicities in the RV data: the strong signal at $P\approx\SI{14.3}{\day}$ and the transiting planet candidate at $P\approx\SI{2.62}{\day}$.
As a result, the basis for our model comparison is a `two-signal model'. Moreover, in \autoref{subsec:activity} we determined the stellar rotation period to be \SI{122}{\day}, {which is recognisable as a peak in the periodogram of the RV data (\autoref{fig:gls_rv}), but not significantly in terms of FAP}. We took this into consideration for the modelling by testing whether an additional {gaussian process (GP)} term that is optimised to mitigate stellar activity signals can improve the fit (referred to as  `three-signal models').

Based on the results from Sects.~\ref{subsec:contamination} and~\ref{subsec:significance_period},
we could assume that the \twoday periodicity is indeed due to a true transiting planet. Therefore, we fixed the period and {time-of-transit} centre for the first model component to the values from the transit-only modelling (\autoref{subsec:transit_only}). This choice is justified because the precision of the transiting planet candidate's {ephemerides} as determined from the photometry is much higher than what could be achieved from the RV data. To investigate the eccentricity of the signal, we tested a sinusoid against a Keplerian model for the transiting planet. Thereby, the eccentricity was parameterised by $\mathcal{S}_1 = \sqrt{e}\sin\omega$ and $\mathcal{S}_2 = \sqrt{e}\cos\omega$ with uniform priors between \num{-1} and \num{1} \citep{Espinoza.2019}. The {prior of the} RV amplitude of the signal was set {uniformly} between \SIrange{0}{50}{\meter\per\second}.

For the \fourteenday signal, we tested a sinusoidal or Keplerian model in the same manner. The period {prior} was set {uniformly} between \SIrange{13.98}{14.71}{\day}, following the analysis with \texttt{AliasFinder} in \autoref{subsubsec:aliasing}, and the {time-of-transit} centre was chosen uniformly between the first epoch of the RV data, \SI{2459061.0}{\day}, and \SI{2459081.0}{\day} to avoid a multi-modal distribution of the posterior.

We investigated whether the RVs are affected by stellar activity by adding a GP component whose {prior on the} rotation period, $P_\text{GP, rv}$, was set {uniformly} between \SIrange{100}{150}{\day} to cover the period determined from the photometry. Our GP kernel was the sum of two simple harmonic oscillators kernels {\citep{ForemanMackey.2017}} as described in \cite{Kossakowski.2021} and hereafter called dSHO-GP ($\equiv$ double simple harmonic oscillator).  The {prior on the} standard deviation, $\sigma_\text{GP,rv}$, of the GP model was specified to be uniform between \SIrange{0}{50}{\meter\per\second} following the Keplerian models. Moreover, we used a uniform prior between {0.1} and 1 for the fractional amplitude, $f_\text{GP,rv}$, of the second component with respect to the first, and log-uniform priors between \numrange{1e-1}{1e4} for the quality factor of the secondary component, $Q_\text{0,GP,rv}$, and the difference compared to the first component, $dQ_\text{GP,rv}$, respectively. For the instrumental parameters of CARMENES, we used uniform priors between \SIrange{-100}{100}{\meter\per\second} for the offset and \SIrange[range-phrase={ \text{to} }]{0}{100}{\meter\per\second} for the jitter.

\paragraph{Results.}
In \autoref{tab:rv_evidence}, we show the Bayesian log evidence for the models that combine the two signals from the periodogram and the stellar activity, as described above. The highest Bayesian log-evidence was found for the model considering sinusoidal components for the \twoday and \fourteenday periods in combination with the GP that accounts for stellar activity. The difference in log-evidence compared to a completely flat model, which means considering only the RV offset and jitter, is  $|\Delta \ln{\mathcal{Z}}|=6$. Following \cite{Trotta.2008}, {we thus assumed the three-signal model}
to be significantly better ($|\Delta \ln\mathcal Z| > 5$).

In comparison with the two-signal models, the models that account for stellar activity are only moderately to almost significantly favoured ($|\Delta \ln\mathcal{Z}| > 2.5 $).
The reason for this is probably the low activity amplitude of \SI{\sim3}{\meter\per\second} {combined with the fact that} only roughly three periods {were covered by the RV observations} ({$\sim350$-day} baseline compared to a period of \SI{\sim120}{\day}). Nonetheless, considering that even small influences from stellar activity can affect the planetary parameters \citep[e.g.][]{Stock.2020b}, and that even strong activity signals do not have to be evident in the periodogram \citep{Nava.2020}, we proceeded with the models that include the GP.

Of these models, those that consider eccentric orbits for one of the two signals are at best indistinguishable $(|\Delta \ln\mathcal{Z}| {<} 1 $) from the model with the highest log evidence, which considers only circular orbits. It can therefore be assumed that the two signals have a low eccentricity, if any. For such {low-eccentricity} orbits, however, the value is mainly determined by the large error bars and the phase coverage of our RV measurements \citep{Hara.2019}{. This ambiguity} is reflected in the indistinguishability of the models and the unconstrained posteriors of the eccentricities ($e_\text{2.6 d} = 0.28 \pm 0.23$; $e_\text{14.3 d} = 0.20 \pm 0.20$). For the {transiting} planet, also considering its short period, it is therefore justified to assume a circular orbit in our further modelling \citep{vanEylen.2019}. Since we do not know the nature of the \fourteenday signal, we proceeded with it in the same way in order to be consistent, and chose the model considering two circular signals for the joint fit.

To exclude the possibility that the choice of our model significantly influences the parameters of the transit planet candidate, we compared the fitted semi-amplitudes and the resulting minimum masses for the different models (\autoref{fig:masscomp}). Additionally, we performed a fit corresponding to the \textbf{2P$_\text{(2.6 d, 14.3 d)}$ + dSHO-GP$_\text{(120 d)}$} model, but replacing the period prior of the \fourteenday signal with a prior considering the \fifteenday alias. All models agree within the interquartile range and show no significant differences. Yet, choosing the \fifteenday alias instead of the \fourteenday period results in a slightly higher planet mass, as is the case for most of the other models. However, those higher masses are also generally accompanied by larger errors.

\subsection{Joint modelling}

\label{subsec:joint_modelling}
\begin{figure*}
    \centering
    \includegraphics[]{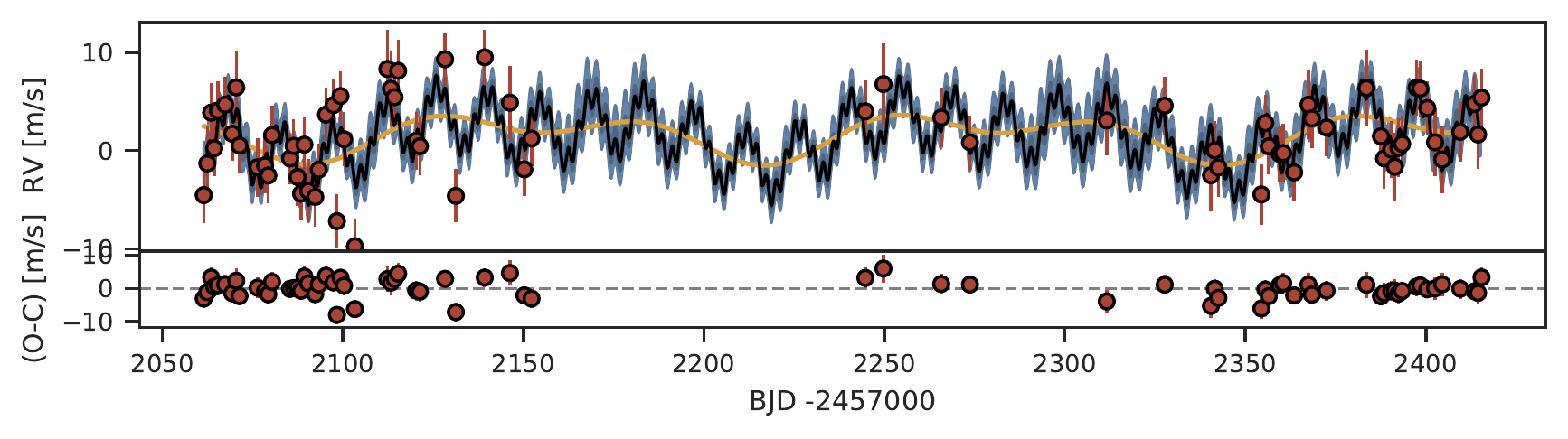}\\
    \includegraphics[]{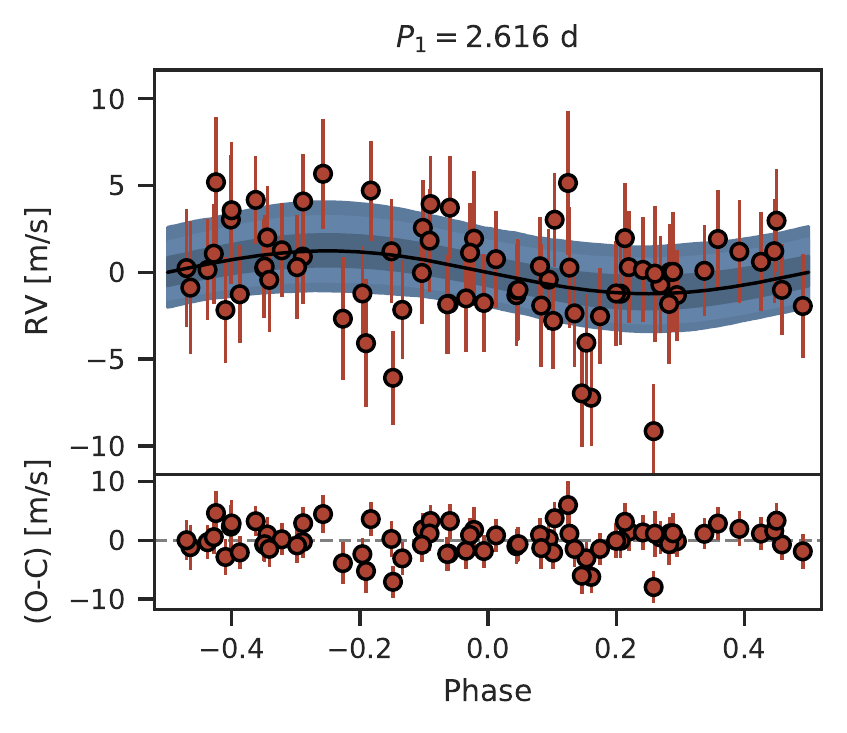}
    \includegraphics[]{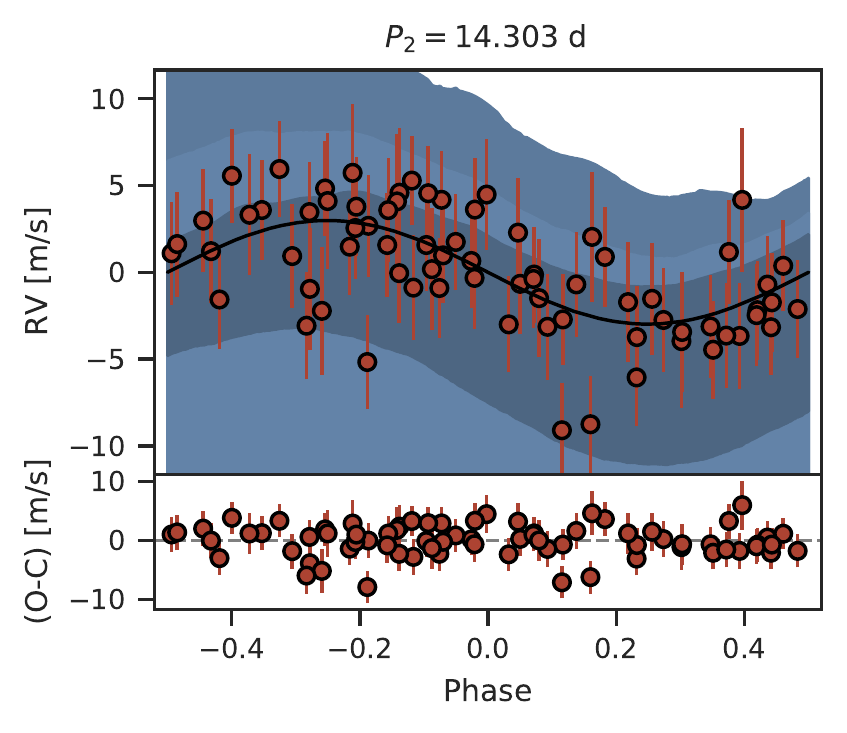}
    \caption{Results for the CARMENES RV from the joint fit with the transits. The black lines show the median of \num{10000} samples from the posterior and the blue shaded areas denote the \SI{68}{\percent}, \SI{95}{\percent,} and \SI{99}{\percent} credibility intervals, respectively. The orange line shows the GP model. Error bars of the measurements include the instrumental jitter added in quadrature. The residuals after subtracting the median models are shown in the lower panels of each plot.
        \textit{Top:} RVs over time. \textit{Bottom:} RVs phase-folded to the periods of the transiting planet (\textit{left}) and the \SI{14.3}{\day} signal (\textit{right}).}
    \label{fig:joint_fit_rv}
\end{figure*}
\begin{figure*}
    \centering
    \includegraphics[]{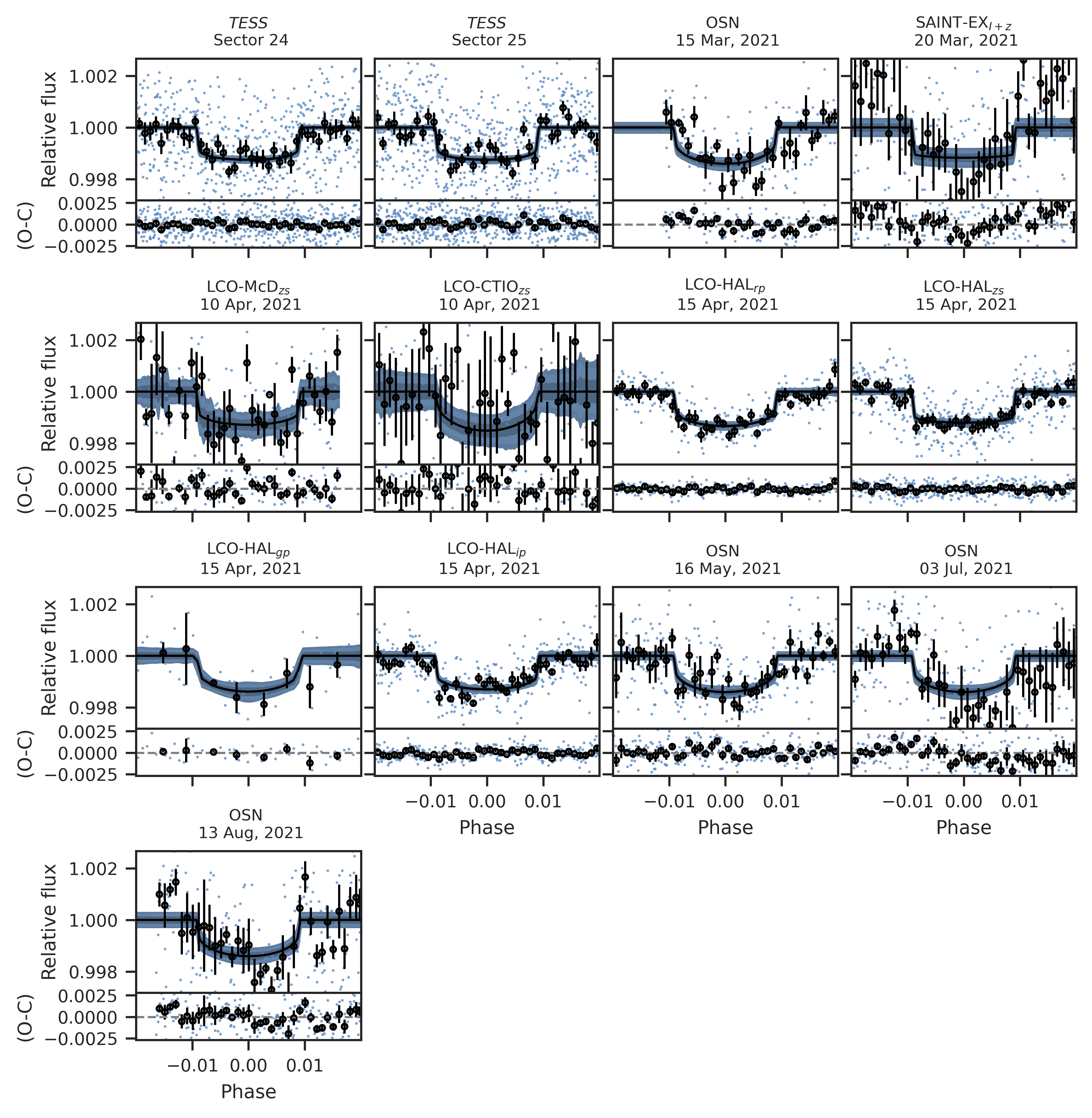}
    \caption{Results from the joint fit for the transit observations. The black lines represent the median of \num{10000} samples from the posterior phase-folded to the period of the transiting planet. Credibility intervals of \SI{68}{\percent}, \SI{95}{\percent,} and \SI{99}{\percent} are displayed by the blue shaded areas. The black points show the data binned to \num{0.001} in phase, and the measurements that were used for the fit are denoted by the blue dots. As for the RVs, the residuals after subtracting the median model are shown in the lower panel of each plot.}
    \label{fig:joint_fit_transit}
\end{figure*}

The highest information content is provided by the combination of the transit and RV data, which is why we performed a joint fit to derive precise parameters of the transiting planet. Based on our results from the transit- and RV-only analyses, the model consists of a circular orbit for the transiting planet with $P\approx\SI{2.62}{\day}$ fitted to the transit and RV data, in combination with the sinusoidal \fourteenday signal, and the {dSHO-GP representing the} stellar activity, in the RV data only. The priors used for the fit correspond to the combination of the transit- and RV-only priors as described in Sects.~\ref{subsec:transit_only} and~\ref{subsec:rv_only}
and are summarised in \autoref{tab:joint_prior}.

We present the posterior parameters of the {transiting} planet, the \fourteenday signal{,} and the GP in \autoref{tab:planet_posterior}, while the posteriors of the instrumental parameters are shown in \autoref{tab:instr_posterior}. Plots of the final models retrieved from the posteriors are shown in \autoref{fig:joint_fit_rv} for the RVs and \autoref{fig:joint_fit_transit} for the transits.

Given the uncertainty of {\SI{35}{\percent}} in the {RV} semi-amplitude of the transiting planet candidate, we checked whether our choice of the \fourteenday signal as the period underlying the $\sim$15-day aliases had a significant effect on the planetary parameters. In \autoref{app:alt_fit} the results from a joint fit considering the \fifteenday period to be the true period are presented. While the derived RV semi-amplitude for the transiting planet candidate is indeed slightly larger, it is fully consistent with the results presented here. Coincidentally, the higher amplitude in combination with the approximately unchanged uncertainties resulted in a significant measurement ({$\sim 3.4\sigma$}). However, following the analysis in \autoref{subsubsec:aliasing} and \autoref{app:alt_alias}, we were confident that $P\approx\SI{14.3}{\day}$ is the true period and, therefore, we accepted the non-significant amplitude from the corresponding fit.

\begin{table}
    \caption{Median posterior parameters of the transiting planet, the \SI{\sim14.3}{\day} signal, and the GP.}
    \label{tab:planet_posterior}
    \centering
    \begin{tabular}{lcl}
        \hline \hline
        Parameter                                   & Posterior\tablefootmark{(a)}                              & Units                            \\
        \hline
        \noalign{\smallskip}
        \multicolumn{3}{c}{\textit{Stellar density}}                                                                                               \\
        $\rho_\star$                                & {$\num{14.96}^{+\num{0.47}}_{-\num{0.59}}$}               & \si{\gram\per\centi\meter\cubed} \\
        \noalign{\smallskip}
        \multicolumn{3}{c}{\textit{GJ~3929~b}}                                                                                                     \\
        $P_\text{{b}}$                              & {$\num{2.6162745}^{+\num{2.9e-06}}_{-\num{3.0e-06}}$}     & d                                \\
        $t_{0,\text{{b}}}$\tablefootmark{(b)}       & {$\num{2459320.05808}^{+\num{0.00018}}_{-\num{0.00019}}$} & d                                \\
        $r_{1,\text{{b}}}$                          & {$\num{0.405}^{+\num{0.060}}_{-\num{0.046}}$}             & \dots                            \\
        $r_{2,\text{{b}}}$                          & {$\num{0.03348}^{+\num{0.00041}}_{-\num{0.00041}}$}       & \dots                            \\
        $K_\text{{b}}$                              & {$\num{1.23}^{+\num{0.40}}_{-\num{0.43}}$}                & $\mathrm{m\,s^{-1}}$             \\
        \noalign{\smallskip}
        \multicolumn{3}{c}{\textit{14.3 d signal}}                                                                                                 \\
        $P_\text{(14.3 d)}$                         & {$\num{14.303}^{+\num{0.034}}_{-\num{0.035}}$}            & d                                \\
        $t_{0, \text{(14.3 d)}}$\tablefootmark{(b)} & {$\num{2459072.44}^{+\num{0.41}}_{-\num{0.41}}$}          & d                                \\
        $K_\text{(14.3 d)}$                         & {$\num{3.04}^{+\num{0.42}}_{-\num{0.44}}$}                & $\mathrm{m\,s^{-1}}$             \\
        \noalign{\smallskip}
        \multicolumn{3}{c}{\textit{GP parameters}}                                                                                                 \\
        $P_\text{GP, rv}$                           & {$\num{126.5}^{+\num{2.4}}_{-\num{2.5}}$}                 & d                                \\
        $\sigma_\text{GP, rv}$                      & {$\num{3.0}^{+\num{2.4}}_{-\num{1.3}}$}                   & $\mathrm{m\,s^{-1}}$             \\
        {$f_\text{GP, rv}$}                         & {$\num{0.84}^{+\num{0.09}}_{-\num{0.11}}$}                & \dots                            \\
        {$Q_{0, \text{GP, rv}}$}                    & {$\num{1110}^{+\num{2270}}_{-\num{750}}$}                 & \dots                            \\
        {$dQ_\text{GP, rv}$}                        & {$\num{1700}^{+\num{3400}}_{-\num{1300}}$}                & \dots                            \\
        \hline
    \end{tabular}
    \tablefoot{\tablefoottext{a}{Error bars denote the $68\%$ posterior credibility intervals.}
        \tablefoottext{b}{Barycentric Julian date in the barycentric dynamical time standard.}}
\end{table}

\subsection{Stellar rotation period}
\label{subsec:activity}

\begin{figure}
    \centering
    \includegraphics{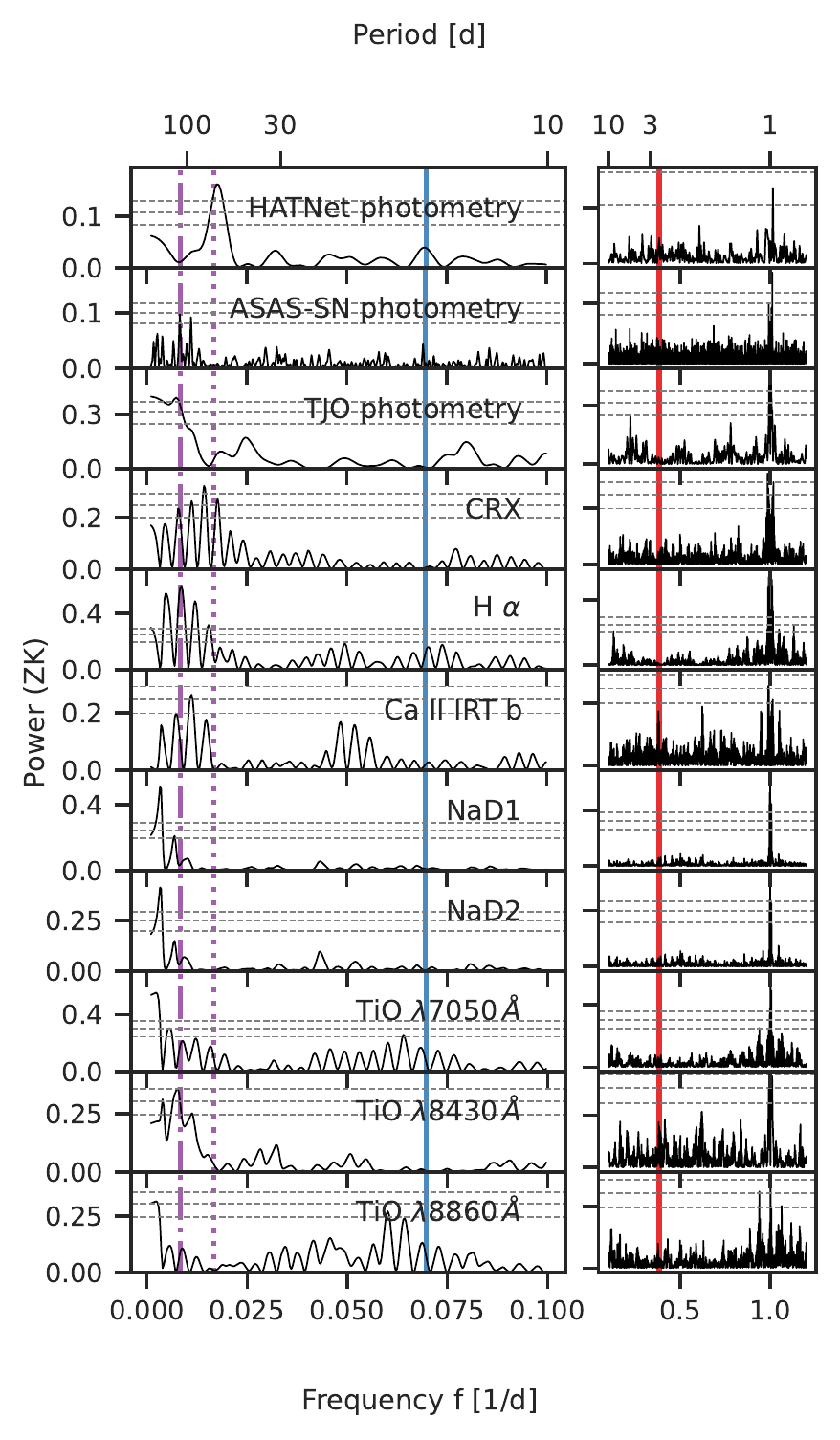}
    \caption{GLS periodograms of photometry and activity indicators. The first three panels show the photometry from HATNet, ASAS-SN, and TJO, and the following panels show the activity indicators derived from CARMENES, which show signals with less than \SI{1}{\percent} FAP. The stellar rotation of $P\approx\SI{122}{\day}$, as determined from the photometry, is indicated by the purple dot-dashed line and its {second} harmonic {($P/2$)} by the purple dotted line. As in \autoref{fig:gls_rv}, the period of the transiting planet is denoted by the red solid line, and the $\sim$\SI{15}{\day} periodicity is marked in blue, respectively. We {normalised} the power using the parameterisation of \cite{Zechmeister.2009} and the 10, 1, and \SI{0.1}{\percent} false alarm probabilities denoted by the horizontal grey dashed lines are calculated using the analytic expression.}
    \label{fig:gls_activity}
\end{figure}

\subsubsection{Activity indicators}

The wide wavelength range of CARMENES {allows us} to compute many indicators that are sensitive to stellar activity. A full list of all activity indicators that are routinely derived from the CARMENES spectra can be found in \citet[spectral indices]{Zechmeister.2018}, \citet[photospheric and chromospheric indices]{Schofer.2019}, and \citet[parameters related to the cross-correlation function]{Lafarga.2020}. For the sake of clarity, we only selected the indicators from the VIS channel that exhibit signals with $\text{FAP} < \SI{1}{\percent}$ {in a GLS periodogram and present them in \autoref{fig:gls_activity}. None of these signals coincide with the period of the transiting planet candidate or the \fourteenday signal. However, all periodograms} show a fairly similar pattern of peaks between \SIrange{50}{300}{\day}. The cause here is also a strong aliasing due to the seasonal observability of GJ~3929 and the {resulting} {strong} sampling frequency of $f_s\approx\SI{1/292}{\per\day}$ (see also \autoref{subsec:rv_only}).

Particularly prominent is the $\text{H} \alpha$ index derived from \texttt{serval}, which shows the strongest peak at a period of \SI{\sim118}{\day}, in combination with its first-order aliases at \SI{\sim82}{\day} and \SI{\sim212}{\day}. Additionally, there is another significant peak of $\sim$65 days, which could be misinterpreted as the {second} harmonic ($\equiv P/2$) of the 118-day period, but it is actually {closer to} its second-order alias.
The {oppositely} signed counterpart of this second order alias produces a significant long-term trend in the data. This is similar to the periodogram of the chromatic index \citep[CRX;][]{Zechmeister.2018}, which is consistent with either an underlying period of {\SI{\sim128}{\day}} that shows aliasing up to third order, or a $\sim$70-day periodicity producing up to second-order aliases. Furthermore, {analogous} patterns can be found for the Ca~{\sc ii} infrared triple~b (IRT~b) index as well as the TiO $\lambda 8430 \AA$ {band}.
{The Na~{\sc i}~D doublet lines and the other two TiO bands are dominated by long-term trends. However, they can also be explained by aliasing of underlying periods of about \SI{113}{\day} (see \autoref{app:activity} for a detailed list of the peaks and corresponding aliases)}.

\subsubsection{Long-term photometry}

We {created} GLS periodograms of the HATNet{, ASAS-SN, and TJO} data (see the first {three} panels of \autoref{fig:gls_activity}). The GLS of the HATNet data shows a highly significant peak at a period of $P\approx\SI{57}{\day}$, which is consistent with the rotation period reported by \cite{Hartman.2011}, {who applied} a variance period-finder using a harmonic series. {However}, the detection of the period was flagged as `questionable' {by the authors} following their visual inspection of the light curve as they did not recognise a clear variability by eye. The ASAS-SN {data}, on the other hand,
show two prominent peaks in the GLS: one at $P\approx\SI{91}{\day}$ and an even more significant one at $P\approx\SI{122}{\day}$. A look at the window function of the data {shows} that these two peaks are generated by aliasing due to a sampling frequency of $f_s\approx1/ \SI{362}{\per\day}$. {The \SI{122}{\day} period is also supported by the TJO data. They show a peak at about \SI{140}{\day}, which  is, due to the short baseline, embedded in a plateau for periods larger than \SI{100}{\day}.}

The photometry and spectroscopic activity indicators thus share a common periodicity of about \SI{\sim120}{\day}, which is about twice the period published by \cite{Hartman.2011} based on the HATNet data alone. However, it is reasonable that $P\approx\SI{120}{\day}$ is the actual rotation period of the star and that HATNet shows the {second} harmonic.

We therefore moved forward and performed a combined fit of the HATNet{, ASAS-SN, and TJO} data using the dSHO-GP model as in \cite{Kossakowski.2021} to determine a precise value for it. In doing so, we used {normally} distributed priors for the instrumental offsets centred around 0 with a standard deviation of 0.1 and log-uniform priors for the instrumental jitter terms between \SIrange{1}{10e6}{ppm}.
For the GP hyperparameters, we used separate instrument priors for the standard deviation, $\sigma_\text{GP,phot}$ (log-uniform between \numrange{1e-8}{1}), the quality factor of the secondary oscillation $Q_0$ and the difference compared to the quality factor of the primary oscillation $dQ_\text{GP,phot}$ (both log uniform between \numrange{0.1}{1e4}), and the fractional amplitude, $f_\text{GP,phot}$ between both (uniform between \numrange{0}{1}). The GP rotation period of $P_\text{GP,phot}$, however, was shared between {all} instruments {with a} uniform {prior} between \SIrange{100}{150}{\day} to avoid the 91-day alias of the ASAS-SN data and the 57-day {second} -order harmonic of the HATNet data. {In this way}, we determined a photometric rotation period of {$P_\text{rot}=\SI{122\pm13}{\day}$}.

We obtained three different measurements of the stellar rotation period: {\SI{126.5\pm2.5}{\day}} from the RV measurements (\autoref{tab:planet_posterior}), {$\sim113$--$132$\,d} from the activity indicators,  and {\SI{122\pm13}{\day}} from the photometry. All three measurements are consistent with each other. {Causes for the differences in the measured periods can be different active latitudes at the times of the measurement, differential rotation, or, in the case of the activity indicators, the differences between photospheric and chromospheric indicators.} Since photometrically determined rotation periods are often considered to be the most reliable and the RV measurement has a likely underestimated uncertainty, as the rotation period of GJ~3929 we adopt the photometric period of {$P_\text{rot}=\SI{122\pm13}{\day}$}, which, {fittingly,} comprises all three measurements the best.

\section{Discussion}
\label{sec:discussion}
\subsection{{GJ~3929}~b}

Our analysis confirms the planetary nature of the transiting planet GJ~3929~b. \autoref{tab:derived_parameters} shows the planetary parameters derived from our joint fit. Its mass and radius of $M_\text{b}=\Mtransiting$ and $R_\text{b}=\Rtransiting$, respectively, put it into the regime of small Earth-sized planets. {This makes GJ~3929~b comparable to other planets with confirmed masses orbiting M stars that were detected by TESS. These include, for example (in order of their detection), L~98--59~b \citep{Kostov.2019, Cloutier.2019, Demangeon.2021}, TOI-270~b {\citep{Gunther.2019, vanEylen.2021}}, GJ~357~b \citep{Luque.2019, Jenkins.2019}, GJ~1252~b \citep{Shporer.2020}, GJ~3473~b \citep{Kemmer.2020}, LHS-1140~c \citep{Ment.2019,LilloBox.2020}, {and}  LHS~1478~b \citep{Soto.2021}.}

\begin{figure}
    \centering
    \includegraphics[]{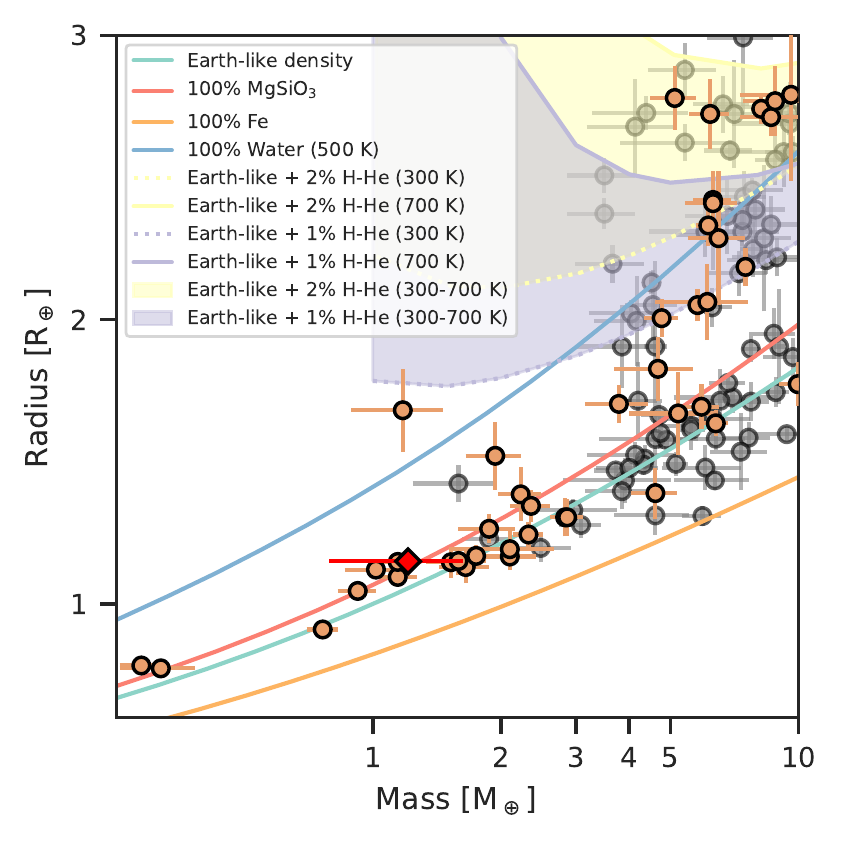}
    \caption{Mass-radius diagram of well-characterised planets with $R<\SI{3}{R_\oplus}$ and $M<\SI{10}{M_\oplus}$. The plot shows the planets from the TEPcat catalogue \citep[][visited on 8 November 2021]{Southworth.2011} with $\Delta M$ and $\Delta R < \SI{30}{\percent}$.
        Planets with host star temperatures $T_\text{eff}<\SI{4000}{K}$ are shown in orange, and planets with hotter hosts are shown in grey. GJ~3929~b is marked with a red diamond. Additionally, theoretical mass-radius relations from \cite{Zeng.2019} are shown for reference.}
    \label{fig:mass_radius}
\end{figure}

{Although the uncertainty in mass allows a wide range of compositions for GJ~3929~b (see \autoref{fig:mass_radius}), its small radius places it below the radius gap for M-dwarf planets \citep{Cloutier.2020c,vanEylen.2021} and, hence, makes a rocky {composition} very likely. The derived mean density of $\rho_\text{b}=\rhotransiting$ is compatible with an MgSiO$_3$-dominated composition. GJ~3929~b thus expands the statistical sample of rocky super-Earths needed to further investigate the properties of the radius gap. For example, as a planet orbiting a mid-type M star, it is {an important contribution} for studies considering the dependence of the gap on the stellar mass or the incident flux, as in \cite{vanEylen.2021}.}

With an orbital period of $P_\text{{b}}=\SI{2.62}{\day}${,} {GJ~3929~b} receives {\num{17.5\pm1.3}} times the solar flux on Earth, which corresponds to an equilibrium temperature of {$T_\text{eq}=\SI{568.8\pm9.4}{\kelvin}$} (assuming zero Bond albedo). In combination with the host star brightness {($J = \SI{8.694}{mag}$)}, this results in a transmission spectroscopy metric \citep[TSM;][]{Kempton.2018} of {\num{25.0\pm13.2}}. GJ~3929~b is thus above the threshold of $\text{TSM} > 10$ determined by \cite{Kempton.2018} and {slightly {larger than} GJ~357~b \citep[$\text{TSM}=  23.4$;][]{Luque.2019}, which is considered one of the prime targets for atmospheric follow-up observations of rocky exoplanets with the upcoming JWST \citep{Gardner.2006}.}

Although unlikely given the Earth-like radius and consequently location below the radius gap, the uncertainty in the determined density does not {completely} exclude the presence of a {significant} atmosphere. {An} atmosphere with a high mean molecular weight {would be} difficult to probe, however, {as it has been shown for other comparable small M-dwarf planets \citep[e.g.][]{Luque.2019,Bower.2019,Nowak.2020}}, the dominant species of carbon dioxide and water are expected to produce absorption features that are observable with instruments {such as the JWST or the ELT \citep{Gilmozzi.2007}}.
The systematic in-depth atmospheric characterisation of rocky planets such as GJ~3929~b is expected to provide answers to questions regarding the abundance and composition of retained primordial atmospheres or secondary atmospheres formed by outgassing.

\begin{table}
    \addtolength{\tabcolsep}{-5pt}
    \caption{Derived planet parameters for GJ~3929~b {and the planet candidate.}}
    \label{tab:derived_parameters}
    \begin{tabular}{lccl}
        \hline \hline
        Parameter                                   & {Posterior P$_\text{b}$\tablefootmark{(a)}}         & {Posterior P$_\text{(14.3 d)}$\tablefootmark{(a)}} & Units                            \\
        \hline
        \noalign{\smallskip}
        \multicolumn{4}{c}{\textit{Derived transit parameters}}                                                                                                                                   \\
        \noalign{\smallskip}
        $p = R_{\rm p}/R_\star$                     & {$\num{0.03348}^{+\num{0.00041}}_{-\num{0.00041}}$} & \dots                                              & \dots                            \\
        $b = (a_{\rm p}/R_\star)\cos i_{\rm p}$     & {$\num{0.108}^{+\num{0.089}}_{-\num{0.069}}$}       & \dots                                              & \dots                            \\
        $a_{\rm p}/R_\star$                         & {$\num{17.56}^{+\num{0.18}}_{-\num{0.24}}$}         & \dots                                              & \dots                            \\
        $i_{\rm p}$                                 & {$\num{89.65}^{+\num{0.23}}_{-\num{0.3}}$}          & \dots                                              & deg                              \\
        \noalign{\smallskip}
        \multicolumn{4}{c}{\textit{Derived physical parameters\tablefootmark{({b})}}}                                                                                                             \\
        \noalign{\smallskip}
        $M_{\rm p}$                                 & {$\num{1.21}^{+\num{0.40}}_{-\num{0.42}}$}          & \dots                                              & $M_\oplus$                       \\
        $M_{\rm p}\sin i$                           & {$\num{1.21}^{+\num{0.40}}_{-\num{0.42}}$}          & {$\num{5.27}^{+\num{0.74}}_{-\num{0.76}}$}         & $M_\oplus$                       \\
        $R_{\rm p}$                                 & {$\num{1.150}^{+\num{0.040}}_{-\num{0.039}}$}       & \dots                                              & $R_\oplus$                       \\
        $\rho_{\rm p}$                              & {$\num{4.4}^{+\num{1.6}}_{-\num{1.6}}$}             & \dots                                              & \si{\gram\per\centi\meter\cubed} \\
        $g_{\rm p}$                                 & {$\num{9.0}^{+\num{3.1}}_{-\num{3.1}}$}             & \dots                                              & \si{\meter\per\second\squared}   \\
        $a_{\rm p}$                                 & {$\num{0.02569}^{+\num{0.00088}}_{-\num{0.00088}}$} & {$\num{0.078}^{+\num{0.0011}}_{-\num{0.0012}}$}    & \si{\astronomicalunit}           \\
        $T_\textnormal{eq, p}$\tablefootmark{({c})} & {$\num{568.8}^{+\num{9.4}}_{-\num{9.3}}$}           & {$\num{326.5}^{+\num{7.6}}_{-\num{7.5}}$}          & \si{\kelvin}                     \\
        $S$                                         & {$\num{17.5}^{+\num{1.3}}_{-\num{1.2}}$}            & {$\num{1.900}^{+\num{0.059}}_{-\num{0.055}}$}      & $S_\oplus$                       \\
        {ESM\tablefootmark{(d)}}                    & {$\num{4.81}^{+\num{0.22}}_{-\num{0.21}}$}          & \dots                                              & \dots                            \\
        {TSM\tablefootmark{(d)}}                    & {$\num{25.0}^{+\num{13.2}}_{-\num{6.3}}$}           & \dots                                              & \dots                            \\
        \hline
    \end{tabular}
    \tablefoot{\tablefoottext{a}{Error bars denote the $68\%$ posterior credibility intervals.}
        \tablefoottext{b}{Sampled from normal distributions for stellar mass, radius, and luminosity based on the results from \autoref{sec:stellar_prop}.}
        \tablefoottext{c}{Assuming a zero Bond albedo.}
        \tablefoottext{d}{Emission and transmission spectroscopy metrics \citep{Kempton.2018}}}
\end{table}

\subsection{Planet candidate GJ~3929~[c]}

The strongest signal in the CARMENES RV data is not related to the transiting planet or the stellar rotation. It has a period of $P_\text{{[c]}}=\Prv$ and an RV semi-amplitude of $K_\text{{[c]}}={\SI{3.04\pm0.44}{\meter\per\second}}$. {A stability analysis of the signal using tools such as the stacked Bayesian GLS \citep{Mortier.2015} is not meaningful because the seasonal observability leads to strong aliasing after the first block of observations and thus a strong loss in signal strength due to the splitting peaks.} In combination with the FAP of the signal, which is still higher than \SI{0.1}{\percent} in the non-pre-whitened periodogram, we thus introduce it as a planet candidate, namely GJ~3929~[c]. The mass derived from the joint fit for this potential planet is $M_\text{[c]}\geq\Mrv$, which puts it into the regime of the sub-Neptune-mass planets.

{In a co-planar orbit, such a planet could be transiting, even if only for a very small range  ({$b_{[c]} =  0.33 \pm 0.23$} assuming the inclination of the inner planet). Full transits should show signals comparable to or larger than those of the less massive inner planet. The fact that we do not detect any other potentially transiting signals in the TESS data after subtracting the \twoday planet suggests that there could be shallow grazing transits, if any. However, confirming the detection of such transits would be complicated by the uncertainty of the ephemeris as determined from the RVs -- {$t_\text{transit, m}= \SI[separate-uncertainty=false]{2 459 072.44\pm0.41}{\day} + m* \SI[separate-uncertainty=false]{14.303\pm0.035}{\day}$} --  which makes it furthermore plausible that some transits could fall just inside the data gaps of TESS}. This is important to note, because applying a TLS periodogram to the unbinned HATNet data in the range of \SIrange[range-phrase={ \text{to} }]{0}{40}{\day} gives rise to a {spurious} signal with a signal detection efficiency
of approximately \num{9.75} (i.e. $\text{FAP}<\SI{0.01}{\percent}$) at a period of $P\approx\SI{14.14}{\day}$ {and with a transit depth of \SI{\approx2.2\pm1.8}{ppt} (see \autoref{fig:tlsgls_hatnet})}. Moreover, even though unexpected, the GLS of the unbinned HATNet data shows, besides the strongest peak at the stellar rotation period, a highly significant peak at $P\approx\SI{14.5}{\day}$ and another peak with $\text{FAP}< \SI{1}{\percent}$ at $P\approx\SI{2.63}{\day}$. Any attempt in fitting the transiting planet together with or without the planet candidate to the HATNet data, however, brought up questionable results. {Furthermore, subtracting the GP model from the determination of the stellar rotation makes the signal disappear. The proximity of the signal's period to half of the Moon's cycle in combination with the large scatter of the unbinned HATNet data ($\text{rms} = \SI{6.7}{ppt}$) suggests a questionable origin of the signal, though we cannot rule out a transit scenario. If the planet were indeed transiting, we would expect a transit depth of \SIrange[range-phrase={ \text{to} }]{\sim1}{8}{ppt} for it based on its minimum mass and the corresponding empirical radius distribution in \autoref{fig:mass_radius}.}

\subsection{{Implications for a multi-planet system}}

Planets such as the candidate GJ~3929~[c] {in combination with} GJ~3929~b are {frequently detected} \citep[e.g.][]{Sabotta.2021,Cloutier.2021}. {Moreover, combinations of terrestrial planets and sub-Neptunes are also commonly predicted by population synthesis models based on the core accretion paradigm of planet formation \citep{Emsenhuber.2021, Schlecker.2021, Burn.2021}.} Following the angular momentum deficit stability criterium from \cite{Laskar.2017}, the system would be stable for eccentricities of the outer companion candidate up to \num{0.45}.

\section{Conclusions}
\label{sec:conclusions}

The analysis of the TESS transit observations, in combination with the RV follow-up from CARMENES and transit follow-up from SAINT-EX and LCOGT, {confirms the planetary nature of the Earth-sized, short-period planet, GJ~3929~b}. Along with the brightness of its M-dwarf host,
{its high equilibrium temperature} makes GJ~3929~b a prime target for atmospheric follow-up with the upcoming generation of {facilities}, such as the JWST, which will provide unique insight into the composition and, thus, formation and evolution of small and rocky planets.

Moreover, the RV measurements showed evidence for a second sub-Neptunian-mass planet candidate, namely GJ~3929~[c]. Its period is far from the rotation period of the star that we determined from archival photometry, and, therefore, it is not likely linked to stellar activity. {Besides this}, the candidate is promising because we detected a signal in the TLS periodogram of archival photometric HATNet data close to the orbital period determined from the RVs. Yet, additional follow-up is needed to confirm its planetary nature, given that the strong aliasing of the RVs and the time gap with respect to the HATNet data made it difficult to provide an in-depth investigation of the signal.

If the planetary nature of GJ~3929~[c] {can} indeed be proven, the GJ~3929 system would join the growing number of multi-planetary systems with relatively short periods around M-dwarf stars. Of particular interest would be whether GJ~3929~[c] is actually a transiting planet and, thus, whether it would be possible to determine its density.

\begin{acknowledgements}
    CARMENES is an instrument for the Centro Astron\'omico Hispano-Alem\'an de Calar Alto (CAHA, Almer\'a, Spain). CARMENES
    is funded by the German Max-Planck-Gesellschaft (MPG), the Spanish Consejo Superior de Investigaciones Científicas (CSIC),
    the European Union through FEDER/ERF FICTS-2011-02 funds, and the members of the CARMENES Consortium (MaxPlanck-Institut für Astronomie, Instituto de Astrofísica de Andalucía, Landessternwarte Königstuhl, Institut de Ciències de
    l'Espai, Insitut für Astrophysik Göttingen, Universidad Complutense de Madrid, Thüringer Landessternwarte Tautenburg,
    Instituto de Astrofísica de Canarias, Hamburger Sternwarte, Centro de Astrobiología and Centro Astronómico HispanoAlemán), with additional contributions by the Spanish Ministry of Economy, the German Science Foundation through the
    Major Research Instrumentation Programme and DFG Research Unit FOR2544 “Blue Planets around Red Stars”, the Klaus
    Tschira Stiftung, the states of Baden-Württemberg and Niedersachsen, and by the Junta de Andalucía.

    Funding for the TESS mission is provided by NASA's Science Mission directorate.
    We acknowledge the use of public TESS Alert data from pipelines at the TESS Science Office and at the TESS Science Processing Operations Center.
    This research has made use of the Exoplanet Follow-up Observation Program website, which is operated by the California Institute of Technology, under contract with the National Aeronautics and Space Administration under the Exoplanet Exploration Program.
    Resources supporting this work were provided by the NASA High-End Computing Program through the NASA Advanced Supercomputing Division at Ames Research Center for the production of the SPOC data products.
    This paper includes data collected by the TESS mission, which are publicly available from the Mikulski Archive for Space Telescopes.

    {This work makes use of observations from the LCOGT network. Part of the LCOGT telescope time was granted by NOIRLab through the Mid-Scale Innovations Program (MSIP). MSIP is funded by NSF. This paper is based on observations made with the MuSCAT3 instrument, developed by the Astrobiology Center and under financial supports by JSPS KAKENHI (JP18H05439) and JST PRESTO (JPMJPR1775), at Faulkes Telescope North on Maui, HI, operated by the Las Cumbres Observatory.}

    {This work includes observations carried out at the Observatorio Astron\'omico Nacional on the Sierra de San Pedro M\'artir (OAN-SPM), Baja California, M\'exico.
    We acknowledge financial support from the Agencia Estatal de Investigaci\'on of the Ministerio de Ciencia, Innovaci\'on y Universidades and the ERDF through projects
    PID2019-109522GB-C5[1:4],
    {PID2019-107061GB-C64, PID2019-110689RB-100},
    ESP2017-87676-C5-1-R,
    and the Centre of Excellence ``Severo Ochoa'' and ``Mar\'ia de Maeztu'' awards to the Instituto de Astrof\'isica de Canarias (CEX2019-000920-S), Instituto de Astrof\'isica de Andaluc\'ia (SEV-2017-0709), and Centro de Astrobiolog\'ia (MDM-2017-0737),
    the Swiss National Science Foundation (PP00P2-163967 and PP00P2-190080),
    the Centre for Space and Habitability of the University of Bern,
    the National Centre for Competence in Research PlanetS, supported by the Swiss National Science Foundation,
    the Deutsche Forschungsgemeinschaft priority program SPP 1992 ``Exploring the Diversity of Extrasolar Planets'' (JE 701/5-1),
    the Excellence Cluster ORIGINS, which is funded by the Deutsche Forschungsgemeinschaft under Germany's Excellence Strategy (EXC-2094 - 390783311),
    NASA (NNX17AG24G),
    JSPS KAKENHI Grant Number JP18H05439,
    JST CREST Grant Number JPMJCR1761,
    the Astrobiology Center of National Institutes of Natural Sciences (NINS) (Grant Number AB031010),
    the UNAM-DGAPA PAPIIT (BG-101321),
    the ``la Caixa'' Foundation (100010434),
    the European Union Horizon 2020 research and innovation programme under the Marie Sk{\l}odowska-Curie (No. 847648, fellowship code LCF/BQ/PI20/11760023),
    and the Generalitat de Catalunya/CERCA programme.}
    {Data were partly collected with the 90-cm telescope at Observatorio de Sierra Nevada (OSN), operated by the Instituto de Astrof\'isica de Andaluc\'i a (IAA, CSIC). We deeply acknowledge the OSN telescope operators for their very appreciable support.}

    The analysis of this work has made use of a wide variety of public available software packages that are not referenced in the manuscript: {\texttt{AstroImageJ} \citep{Collins.2017}}, \texttt{astropy} \cite{AstropyCollaboration.2018},
    \texttt{scipy} {\citep{Virtanen.2020}}, \texttt{numpy} \citep{Oliphant.2006}, \texttt{matplotlib} \citep{Hunter.2007}, \texttt{tqdm} \citep{daCostaLuis.2019}, \texttt{pandas} \citep{Thepandasdevelopmentteam.2020}, \texttt{seaborn} \citep{Waskom.2020}, \texttt{lightkurve} \citep{LightkurveCollaboration.2018}, \texttt{radvel} \citep{Fulton.2018}, \texttt{batman} \citep{Kreidberg.2015}, \texttt{dynesty} \citep{Speagle.2020}, \texttt{george} \citep{Ambikasaran.2015}, \texttt{celerite} \citep{ForemanMackey.2017}, \texttt{PyFITS} \citep{Barrett.2012}, \texttt{astrobase} \cite{WaqasBhatti.2020}, \texttt{scikit-learn} \citep{Pedregosa.2011}, {\texttt{TAPIR} \citep{Jensen.2013}}, \texttt{Astrometry.net} \citep{Lang.2010}, \texttt{photutils} \citep{Bradley.2021}, \texttt{lmfit} \citep{Newville.2021}, \texttt{tpfplotter} (\url{www.github.com/jlillo/tpfplotter}).
\end{acknowledgements}

\clearpage
\bibliographystyle{aa}
\bibliography{references}

\begin{appendix} 
    \section{Differentiating aliases using posterior samples}
    \label{app:alt_alias}
    The \texttt{AliasFinder} is based on creating synthetic data from the periods, amplitudes, and phases retrieved from an observed periodogram {\citep{Stock.2020c, Stock.2020b}}. One advantage of this {approach} is that it allows us to perform an alias analysis solely from a given set of RV measurements. However, in the presence of {additional signals unrelated to the aliases in question (e.g. additional planets or stellar activity)}, those have to be taken into account by pre-whitening the data to mitigate their influence on the observed periodogram and make it comparable to the synthetic data. {Yet, stellar activity, which often produces only quasi-periodic signals, poses a problem in this respect: the GPs that are commonly used to model such signals are {not} static model components, but they {parametrise} the covariance between the data points. This makes the pre-whitening impossible, because if the signal in question is omitted during pre-whitening, the GP model will be influenced by it and may absorb it from the residuals. Conversely, if the signal in question is taken into account in the pre-whitening (and later reinserted into the data), the GP model implies its presence in the residuals. In both cases, the signals recovered from the residual data do not resemble those of the original data and thus defeat the object of \texttt{AliasFinder}.}

        {Bayesian modelling approaches, {such as} the nested sampling used in our analysis, however, offer a direct solution to this issue: the results from the posterior can also be adopted to generate the synthetic RV models used to create the comparison periodograms. In this way each model can include all required  components of the fit and thus make the resulting periodograms directly comparable with the observed periodogram. Pre-whitening is no longer necessary in this case.}

    The procedure is as follows: For each {possible alias} period that is going to be investigated, a fit has to be performed. {Thereby, the period of the fitted alias signal needs to be reasonably constrained, such that other aliases are excluded. Furthermore, the fit should consider all other signals of interest.} Then, the synthetic RVs {can be} created using the {solutions from the} individual posterior samples {of} the fit results. For each  sample that is drawn, the RV model is calculated on the time stamps of the observations {and the uncertainties of the original measurements are adopted --- {analogously} to the method of the \texttt{AliasFinder} and \cite{Dawson.2010}}. {These} model RVs, however, do not include any {noise} and would therefore result in highly significant peaks in the periodogram for each considered period. A good measure of the noise is the rms of the residuals after subtracting it from the observed data. {As an analogy} to the jitter determination in the \texttt{AliasFinder}, one can therefore add{ white noise} to the synthetic models that are drawn from a normal distribution {and} follow the residual rms. {The evaluation is then analogous to the \texttt{AliasFinder}. After calculating the GLS periodogram for all synthetic RV datasets, the median GLS and its confidence {intervals}, as well as the phases {of the peaks,} can be determined and compared to the observed GLS and its phases.}

\begin{figure}[!ht]
        \centering
        \includegraphics{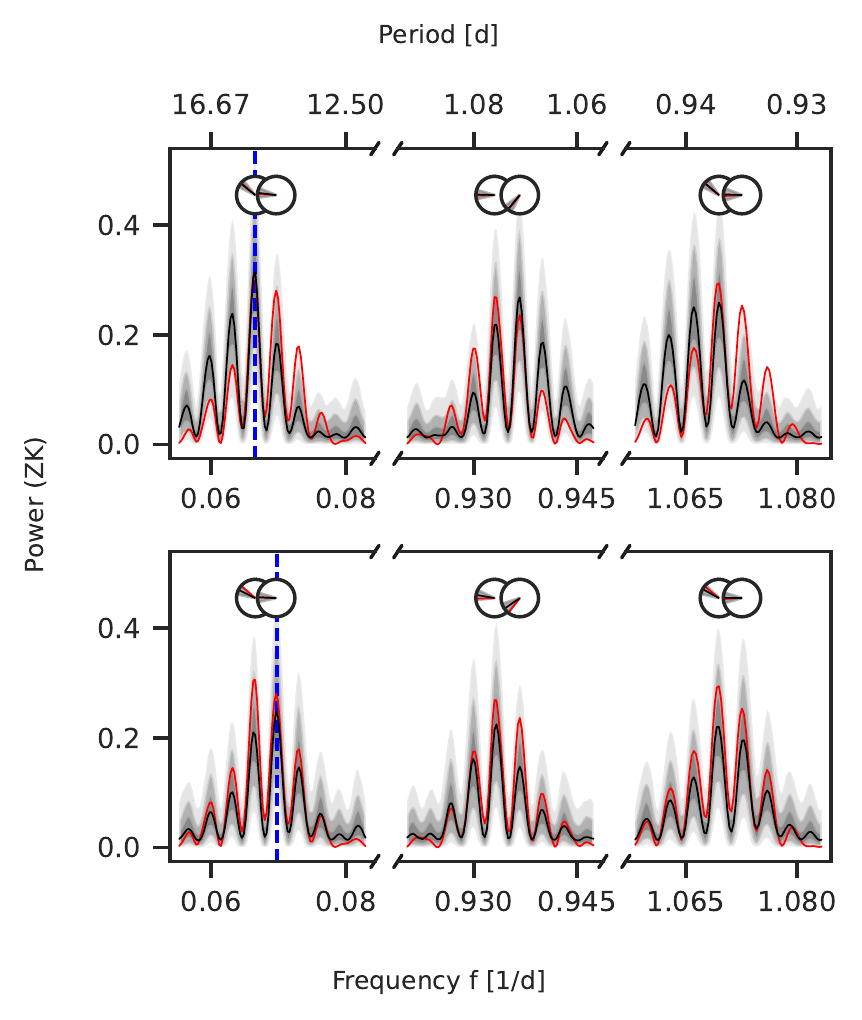}
        \caption{Alias test for the \SI{\sim15}{\day} and \SI{\sim14.3}{\day} periods using the posterior samples from the RV-only fits. We took \num{5000} posterior samples of the second component from the $\bf2P+dSHO-GP_\text{120 d}$ models to produce synthetic periodograms (black lines), which can be compared with the periodogram of the observed data (red lines). The results for the model considering an \SI{\sim15}{\day} signal are shown in the first row, and the results for the \SI{\sim14.3}{\day} signal are in the second row, with each period indicated by a vertical blue dashed line, respectively. Black lines depict the median of the samples for each simulation, and the grey shaded areas are the 50, 90, and \SI{99}{\percent} confidence intervals. Furthermore, the phases of the peaks as measured from the GLS are displayed in the circles, following the same colour scheme (the grey shades denote the standard deviations of the simulated peaks).}
        \label{fig:aliasposterior}
    \end{figure}

    In \autoref{fig:aliasposterior}, we present the results from the $\bf2P+dSHO-GP_\text{120 d}$ models as described in \autoref{subsec:rv_only}, where the second period was either constrained to the \fourteenday or \fifteenday period. The resulting synthetic periodograms are consistent with the results using the \texttt{AliasFinder} and also confirm that considering the \fourteenday period results in a better match to the observed periodogram.

    \clearpage

    \section{Additional figures {and tables}}
    \begin{figure}[!ht]
        \centering
        \includegraphics{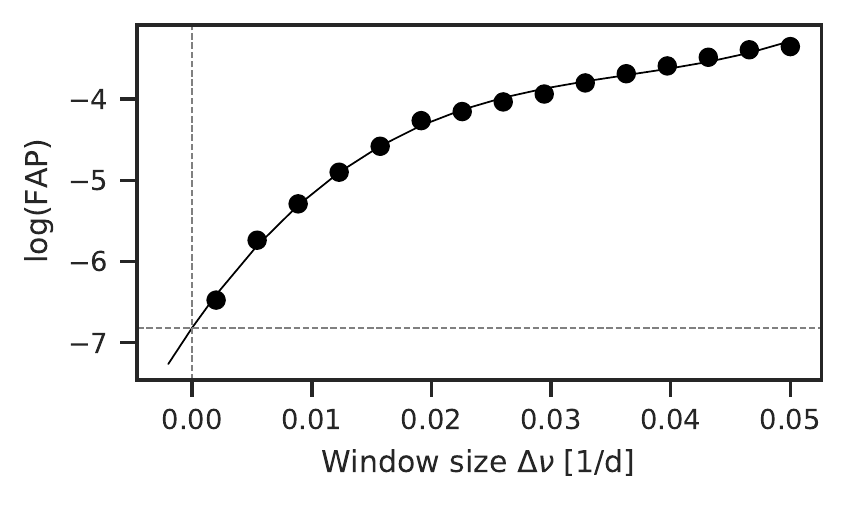}
        \caption{{Determining the FAP for the signal of the transiting planet in the RVs. For each window size, the FAP was calculated comparing the combined power of the highest peaks appearing around the \twoday period of the transiting planet candidate and the \oneday alias from \num{50000} permutations with the combined power of the signals in the original GLS.} The black line shows a third-order polynomial fit to the data, which is extrapolated to zero to determine the FAP.}
        \label{fig:fap}
    \end{figure}

    \begin{figure}[!ht]
        \centering
        \includegraphics{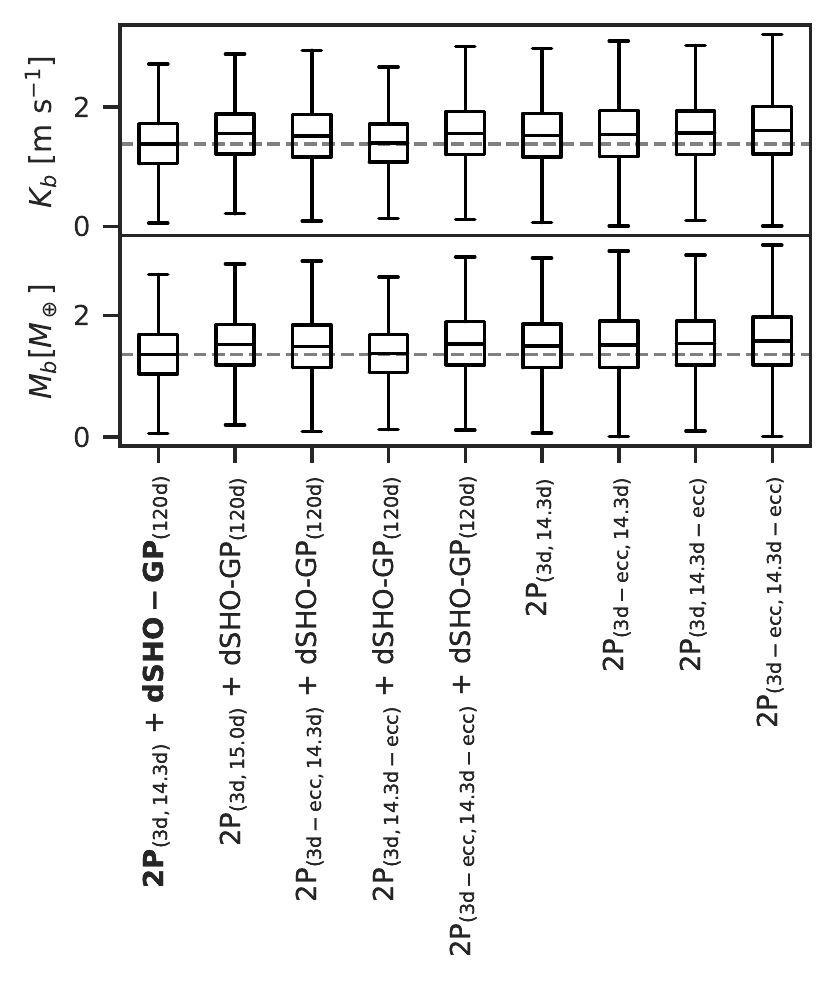}
        \caption{Comparison of the amplitudes and resulting minimum masses for the different models considered. The box plots show the posterior distribution from the model comparison presented in the RV-only analysis (\autoref{subsec:rv_only}). The width of each box corresponds to the interquartile range (IQR), and the whiskers mark the first quartile minus $1.5 \times$ the IQR and the third quartile plus $1.5 \times$ the IQR, respectively. {To facilitate the comparison, the grey dashed line shows the median posterior value of our selected model.}}
        \label{fig:masscomp}
    \end{figure}

    \begin{figure}[!ht]
        \centering
        \includegraphics{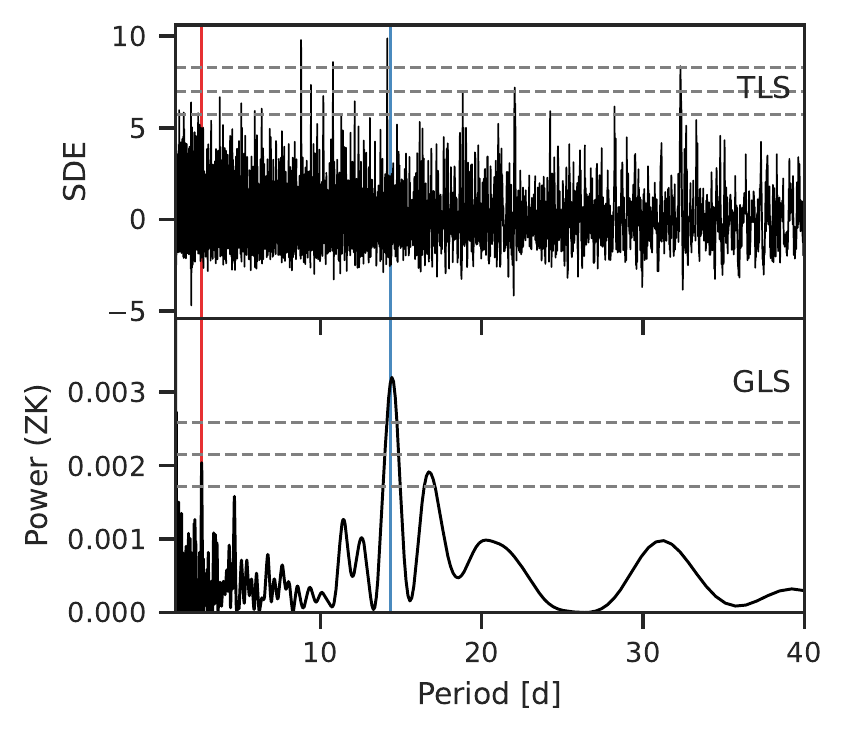}
        \caption{TLS and GLS periodgrams of the unbinned HATNet data. The periods of GJ~3929~b and the candidate GJ~3929~[c] are marked by the vertical red and blue lines, respectively.}
        \label{fig:tlsgls_hatnet}
    \end{figure}

    \begin{table}[!hb]
        \centering
        \caption{Photometry of GJ~3929.}
        \label{tab:photometric_passbands}
        \begin{tabular}{lS[table-format=2.4(4)]c}
            \hline \hline
            Passband & {Brightness [mag]} & Source           \\
            \hline
            $B$      & 14.33\pm0.01       & UCAC4            \\
            $g$      & 13.47\pm0.01       & UCAC4            \\
            $BP$     & 12.9667\pm0.0030   & \emph{Gaia} EDR3 \\
            $V$      & 12.675\pm0.020     & UCAC4            \\
            $r$      & 12.1\pm0.03        & UCAC4            \\
            $G$      & 11.5066\pm0.0028   & \emph{Gaia} EDR3 \\
            $i$      & 10.921\pm0.001     & SDSS9            \\
            $RP$     & 10.3212\pm0.0038   & \emph{Gaia} EDR3 \\
            $J$      & 8.694\pm0.024      & 2MASS            \\
            $H$      & 8.10\pm0.15        & 2MASS            \\
            $K_s$    & 7.869\pm0.020      & 2MASS            \\
            $W1$     & 7.688\pm0.024      & WISE             \\
            $W2$     & 7.540\pm0.020      & WISE             \\
            $W3$     & 7.419\pm0.016      & WISE             \\
            $W4$     & 7.267\pm0.085      & WISE             \\
            \hline
        \end{tabular}
        \tablebib{
            UCAC4: \citet{Zacharias.2013};
            \emph{Gaia} EDR3: \citet{GaiaCollaboration.2021};
            SDSS9: \citet{Ahn.2012};
            2MASS: \citet{Skrutskie.2006};
            WISE: \citet{Cutri.2012}.
        }
    \end{table}

    \clearpage

    \onecolumn
    \section{Priors for \texttt{juliet}}
    \begin{center}
        \begin{longtable}{lllp{0.4\textwidth}}
            \caption{Priors used for \texttt{juliet} in the joint fit of transits and RV.}
            \label{tab:joint_prior}                                                                                                                                                                                        \\
            \hline
            \hline
            \multicolumn{1}{l}{Parameter}                         &
            \multicolumn{1}{l}{Prior}                             &
            \multicolumn{1}{l}{Units}                             &
            \multicolumn{1}{l}{Description}                                                                                                                                                                                \\
            \hline
            \endhead
            \hline \multicolumn{4}{l}{{Continued on next page}}
            \endfoot
            \hline
            \endlastfoot

            \multicolumn{4}{c}{\textit{Stellar parameters}}                                                                                                                                                                \\
            \noalign{\smallskip}
            $\rho_\star$                                          & $\mathcal{N}({13.880, 2})$          & {\si{\gram\per\centi\meter\cubed}} & Stellar density                                                             \\
            \noalign{\smallskip}
            \multicolumn{4}{c}{\textit{Stable components}}                                                                                                                                                                 \\
            \noalign{\smallskip}
            $P_\text{{b}}$                                        & $\mathcal{U}(2.0, 3.0)$             & d                                  & Period of the transiting planet                                             \\
            $t_{0,\text{{b}}}$                                    & $\mathcal{U}(2459319.0, 2459322.0)$ & d                                  & Time of transit centre of the transiting planet                             \\
            $r_{1,\text{{b}}}$                                    & $\mathcal{U}(0, 1)$                 & \dots                              & {Parameterisation} for p and b                                              \\
            $r_{2,\text{{b}}}$                                    & $\mathcal{U}(0, 1)$                 & \dots                              & {Parameterisation} for p and b                                              \\
            $K_\text{{b}}$                                        & $\mathcal{U}(0, 50)$                & $\mathrm{m\,s^{-1}}$               & Radial-velocity semi-amplitude of the transiting planet                     \\
            $\sqrt{e_\text{{b}}}\sin \omega_\text{{b}}$           & $\mathrm{fixed}\,(0)$               & \dots                              & {Parameterisation} for $e$ and $\omega$.                                    \\
            $\sqrt{e_\text{{b}}}\cos \omega_\text{{b}}$           & $\mathrm{fixed}\,(0)$               & \dots                              & {Parameterisation} for $e$ and $\omega$.                                    \\
            \noalign{\smallskip}
            $P_\text{(14.3 d)}$                                   & $\mathcal{U}(13.98, 14.71)$         & d                                  & Period of the second RV signal                                              \\
            $t_\text{0, (14.3 d)}$                                & $\mathcal{U}(2459061.0, 2459081.0)$ & d                                  & Time of transit centre of the second RV signal                              \\
            $K_\text{(14.3 d)}$                                   & $\mathcal{U}(0, 50)$                & $\mathrm{m\,s^{-1}}$               & Radial-velocity semi-amplitude of the second RV signal                      \\
            $\sqrt{e_\text{(14.3 d)}}\sin \omega_\text{(14.3 d)}$ & $\mathrm{fixed}\,(0)$               & \dots                              & {Parameterisation} for $e$ and $\omega$.                                    \\
            $\sqrt{e_\text{(14.3 d)}}\cos \omega_\text{(14.3 d)}$ & $\mathrm{fixed}\,(0)$               & \dots                              & {Parameterisation} for $e$ and $\omega$.                                    \\
            \noalign{\smallskip}
            \multicolumn{4}{c}{\textit{RV GP component}}                                                                                                                                                                   \\
            \noalign{\smallskip}
            $P_\text{GP, rv}$                                     & $\mathcal{U}(100, 150)$             & d                                  & Rotation period of the primary mode                                         \\
            $\sigma_\text{GP, rv}$                                & $\mathcal{U}(0, 10)$                & $\mathrm{m\,s^{-1}}$               & The standard deviation of the GP                                            \\
            $Q_{0, \text{GP, rv}}$                                & $\mathcal{J}(0.1, 10000)$           & \dots                              & Quality factor of the secondary mode                                        \\
            $dQ_\text{GP, rv}$                                    & $\mathcal{J}(0.1, 10000)$           & \dots                              & Difference between the quality factors of the primary and secondary modes   \\
            $f_\text{GP, rv}$                                     & $\mathcal{U}(0.1, 1.0)$             & \dots                              & Fractional amplitude of the secondary mode                                  \\
            \multicolumn{4}{c}{\textit{Instrument parameters CARMENES}}                                                                                                                                                    \\
            \noalign{\smallskip}
            $\mu$                                                 & $\mathcal{U}(-100, 100)$            & $\mathrm{m\,s^{-1}}$               & Instrumental offset                                                         \\
            $\sigma$                                              & $\mathcal{U}(0, 100)$               & $\mathrm{m\,s^{-1}}$               & Jitter term                                                                 \\
            \noalign{\smallskip}
            \multicolumn{4}{c}{\textit{Instrument parameters} TESS}                                                                                                                                                        \\
            $q_{1}$                                               & $\mathcal{U}(0.0, 1.0)$             & \dots                              & Quadratic limb-darkening parameterisation, shared between Sectors 24 and 25 \\
            $q_{2}$                                               & $\mathcal{U}(0.0, 1.0)$             & \dots                              & Quadratic limb-darkening parameterisation, shared between Sectors 24 and 25 \\
            mdilution                                             & $\mathrm{fixed}\,(1)$               & \dots                              & Dilution factor                                                             \\
            mflux                                                 & $\mathcal{N}(0.0, 0.01)$            & \dots                              & Instrumental offset                                                         \\
            $\sigma$                                              & $\mathcal{U}(1, 500)$               & ppm                                & Jitter term                                                                 \\
            \noalign{\smallskip}
            \multicolumn{4}{c}{\textit{Instrument parameters} SAINT-EX, LCOGT{, OSN}}                                                                                                                                      \\
            $q_{1}$                                               & $\mathcal{U}(0, 1)$                 & \dots                              & Linear limb-darkening parameterisation                                      \\
            mdilution                                             & $\mathrm{fixed}\,(1)$               & \dots                              & Dilution factor                                                             \\
            mflux                                                 & $\mathcal{N}(0, 0.1)$               & \dots                              & Instrumental offset                                                         \\
            $\sigma$                                              & $\mathcal{U}(1, 5000)$              & ppm                                & Jitter term                                                                 \\
            \noalign{\smallskip}
            \multicolumn{4}{c}{\textit{De-trending parameters} LCOGT CTIO$_{z_{{s}}'}$}                                                                                                                                    \\
            $\theta_0$                                            & $\mathcal{N}(0, \num{1e-08})$       & \dots                              & Linear de-trending with the comparison ensemble counts                      \\
            $\theta_1$                                            & $\mathcal{N}(0, 0.001)$             & \dots                              & Linear de-trending with target FWHM                                         \\
            \noalign{\smallskip}
            \multicolumn{4}{c}{\textit{De-trending parameters} LCOGT McD$_{z_{{s}}'}$}                                                                                                                                     \\
            $\theta_0$                                            & $\mathcal{N}(0, \num{1e-08})$       & \dots                              & Linear de-trending with the comparison ensemble counts                      \\
            $\theta_1$                                            & $\mathcal{N}(0, 0.001)$             & \dots                              & Linear de-trending with target FWHM                                         \\
            \noalign{\smallskip}
            \multicolumn{4}{c}{\textit{De-trending parameters} LCOGT HAL$_{g'}$}                                                                                                                                           \\
            $\theta_0$                                            & $\mathcal{N}(0, \num{1e-08})$       & \dots                              & Linear de-trending with the comparison ensemble counts                      \\
            $\theta_1$                                            & $\mathcal{N}(0, 0.1)$               & \dots                              & Linear de-trending with the BJD timestamps                                  \\
            \noalign{\smallskip}
            \multicolumn{4}{c}{\textit{De-trending parameters} LCOGT HAL$_{r'}$}                                                                                                                                           \\
            $\theta_0$                                            & $\mathcal{N}(0, \num{1e-08})$       & \dots                              & Linear de-trending with comparison ensemble counts                          \\
            $\theta_1$                                            & $\mathcal{N}(0, 0.1)$               & \dots                              & Linear de-trending with the BJD timestamps                                  \\
            \noalign{\smallskip}
            \multicolumn{4}{c}{\textit{De-trending parameters} LCOGT HAL$_{i'}$}                                                                                                                                           \\
            $\theta_0$                                            & $\mathcal{N}(0, 0.0001)$            & \dots                              & Linear de-trending with the sky background                                  \\
            $\theta_1$                                            & $\mathcal{N}(0, 0.1)$               & \dots                              & Linear de-trending with the BJD timestamps                                  \\
            \noalign{\smallskip}
            \multicolumn{4}{c}{\textit{De-trending parameters} LCOGT HAL$_{z_{{s}}'}$}                                                                                                                                     \\
            $\theta_0$                                            & $\mathcal{N}(0, 0.001)$             & \dots                              & Linear de-trending with the target FWHM                                     \\
            $\theta_1$                                            & $\mathcal{N}(0, 0.1)$               & \dots                              & Linear de-trending with the BJD timestamps                                  \\
            \hline
        \end{longtable}
        \tablefoot{The prior labels $\mathcal{U}$, $\mathcal{J}$, and $\mathcal{N}$ represent uniform, log-uniform, and normal distributions, respectively.}
    \end{center}

    \clearpage

    \section{{Instrumental posteriors of the joint fit}}

     {\setlength{\extrarowheight}{4.5pt}
      \begin{table*}[!ht]
          \centering
          \caption{Posteriors of the joint fit for the different instrumental parameters.}
          \label{tab:instr_posterior}
          \begin{tabular}[t]{lcl}
              \hline
              \hline
              Parameter  & Posterior\tablefootmark{(a)}                        & Units                  \\
              \hline
              \noalign{\smallskip}
              \multicolumn{3}{c}{CARMENES}                                                              \\
              $\mu$      & $\num{1.63}^{+\num{0.37}}_{-\num{0.35}}$            & \si{\meter\per\second} \\
              $\sigma$   & $\num{2.28}^{+\num{0.3}}_{-\num{0.27}}$             & ppm                    \\
              \noalign{\smallskip}
              \multicolumn{3}{c}{TESS Sector 24}                                                        \\
              mflux      & $\num{-0.000126}^{+\num{2.8e-05}}_{-\num{2.9e-05}}$ & \dots                  \\
              $\sigma$   & $\num{4.5}^{+\num{8.2}}_{-\num{2.7}}$               & ppm                    \\
              \noalign{\smallskip}
              \multicolumn{3}{c}{TESS Sector 25}                                                        \\
              $q_1$      & $\num{0.154}^{+\num{0.085}}_{-\num{0.075}}$         & \dots                  \\
              $q_2$      & $\num{0.33}^{+\num{0.12}}_{-\num{0.12}}$            & \dots                  \\
              mflux      & $\num{-0.000102}^{+\num{3e-05}}_{-\num{2.9e-05}}$   & \dots                  \\
              $\sigma$   & $\num{4.9}^{+\num{8.6}}_{-\num{3.0}}$               & ppm                    \\
              \noalign{\smallskip}
              \multicolumn{3}{c}{SAINT-EX$_{I+z}$ 19 Mar. 2021}                                         \\
              $q_1$      & $\num{0.138}^{+\num{0.093}}_{-\num{0.082}}$         & \dots                  \\
              mflux      & $\num{-8e-05}^{+\num{0.00013}}_{-\num{0.00013}}$    & \dots                  \\
              $\sigma$   & $\num{17}^{+\num{49}}_{-\num{13}}$                  & ppm                    \\
              \noalign{\smallskip}
              \multicolumn{3}{c}{LCO CTIO$_{zs'}$ 10 Apr. 2021}                                         \\
              $q_1$      & $\num{0.8}^{+\num{0.12}}_{-\num{0.15}}$             & \dots                  \\
              mflux      & $\num{-0.0011}^{+\num{0.0027}}_{-\num{0.0029}}$     & \dots                  \\
              $\sigma$   & $\num{3060}^{+\num{170}}_{-\num{160}}$              & ppm                    \\
              $\theta_0$ & $\num{1.9e-09}^{+\num{1.2e-09}}_{-\num{1.3e-09}}$   & \dots                  \\
              $\theta_1$ & $\num{-0.00038}^{+\num{0.00024}}_{-\num{0.00024}}$  & \dots                  \\
              \noalign{\smallskip}
              \multicolumn{3}{c}{LCO McD$_{zs'}$ 10 Apr. 2021}                                          \\
              $q_1$      & $\num{0.4}^{+\num{0.12}}_{-\num{0.13}}$             & \dots                  \\
              mflux      & $\num{-0.0128}^{+\num{0.0057}}_{-\num{0.0052}}$     & \dots                  \\
              $\sigma$   & $\num{1760}^{+\num{130}}_{-\num{110}}$              & ppm                    \\
              $\theta_0$ & $\num{-6.7e-09}^{+\num{3.2e-09}}_{-\num{2.5e-09}}$  & \dots                  \\
              $\theta_1$ & $\num{0.0}^{+\num{0.0002}}_{-\num{0.0002}}$         & \dots                  \\
              \noalign{\smallskip}
              \multicolumn{3}{c}{LCO HAL$_{g'}$ 15 Apr. 2021}                                           \\
              $q_1$      & $\num{0.59}^{+\num{0.17}}_{-\num{0.18}}$            & \dots                  \\
              mflux      & $\num{-0.026}^{+\num{0.011}}_{-\num{0.012}}$        & \dots                  \\
              $\sigma$   & $\num{49}^{+\num{166}}_{-\num{38}}$                 & ppm                    \\
              $\theta_0$ & $\num{-2.1e-09}^{+\num{3.1e-09}}_{-\num{3.8e-09}}$  & \dots                  \\
              $\theta_1$ & $\num{-0.0187}^{+\num{0.0019}}_{-\num{0.0019}}$     & \dots                  \\
              \hline
          \end{tabular}\quad\quad\quad\quad
          \begin{tabular}[t]{lcl}
              \hline
              \hline
              Parameter  & Posterior\tablefootmark{(a)}                        & Units \\
              \hline
              \noalign{\smallskip}
              \multicolumn{3}{c}{LCO HAL$_{i'}$ 15 Apr. 2021}                          \\
              $q_1$      & $\num{0.428}^{+\num{0.081}}_{-\num{0.088}}$         & \dots \\
              mflux      & $\num{0.00422}^{+\num{0.001}}_{-\num{0.00097}}$     & \dots \\
              $\sigma$   & $\num{519}^{+\num{31}}_{-\num{29}}$                 & ppm   \\
              $\theta_0$ & $\num{4.17e-05}^{+\num{6.4e-06}}_{-\num{6.8e-06}}$  & \dots \\
              $\theta_1$ & $\num{0.0004}^{+\num{0.0015}}_{-\num{0.0015}}$      & \dots \\
              \noalign{\smallskip}
              \multicolumn{3}{c}{LCO HAL$_{r'}$ 15 Apr. 2021}                          \\
              $q_1$      & $\num{0.493}^{+\num{0.093}}_{-\num{0.099}}$         & \dots \\
              mflux      & $\num{-0.0279}^{+\num{0.0046}}_{-\num{0.0035}}$     & \dots \\
              $\sigma$   & $\num{304}^{+\num{37}}_{-\num{37}}$                 & ppm   \\
              $\theta_0$ & $\num{-1.64e-09}^{+\num{2.2e-10}}_{-\num{1.8e-10}}$ & \dots \\
              $\theta_1$ & $\num{0.00513}^{+\num{0.00068}}_{-\num{0.00066}}$   & \dots \\
              \noalign{\smallskip}
              \multicolumn{3}{c}{LCO HAL$_{zs'}$ 15 Apr. 2021}                         \\
              $q_1$      & $\num{0.21}^{+\num{0.11}}_{-\num{0.11}}$            & \dots \\
              mflux      & $\num{0.00394}^{+\num{0.00095}}_{-\num{0.00096}}$   & \dots \\
              $\sigma$   & $\num{433}^{+\num{34}}_{-\num{35}}$                 & ppm   \\
              $\theta_0$ & $\num{-0.000508}^{+\num{6.4e-05}}_{-\num{7.1e-05}}$ & \dots \\
              $\theta_1$ & $\num{0.01104}^{+\num{0.00062}}_{-\num{0.00061}}$   & \dots \\
              \noalign{\smallskip}
              \multicolumn{3}{c}{OSN 14 Mar. 2021}                                     \\
              $q_1$      & $\num{0.63}^{+\num{0.12}}_{-\num{0.12}}$            & \dots \\
              mflux      & $\num{-0.000596}^{+\num{7.5e-05}}_{-\num{7.2e-05}}$ & \dots \\
              $\sigma$   & $\num{254}^{+\num{108}}_{-\num{85}}$                & ppm   \\
              \noalign{\smallskip}
              \multicolumn{3}{c}{OSN 16 May 2021}                                      \\
              mflux      & $\num{-0.00051}^{+\num{5.7e-05}}_{-\num{5.8e-05}}$  & \dots \\
              $\sigma$   & $\num{19}^{+\num{52}}_{-\num{15}}$                  & ppm   \\
              \noalign{\smallskip}
              \multicolumn{3}{c}{OSN 2 July 2021}                                      \\
              mflux      & $\num{-0.000107}^{+\num{8e-05}}_{-\num{7.9e-05}}$   & \dots \\
              $\sigma$   & $\num{111.0}^{+\num{131.0}}_{-\num{67.0}}$          & ppm   \\
              \noalign{\smallskip}
              \multicolumn{3}{c}{OSN 13 Aug. 2021}                                     \\
              mflux      & $\num{-0.00076}^{+\num{0.00011}}_{-\num{0.00011}}$  & \dots \\
              $\sigma$   & $\num{12.7}^{+\num{45.3}}_{-\num{9.5}}$             & ppm   \\
              \hline
          \end{tabular}

          \tablefoot{\tablefoottext{a}{Error bars denote the $68\%$ posterior credibility intervals.}}
      \end{table*}}

    \clearpage
    \twocolumn
    \section{Alternative joint fit considering $P_2\sim\SI{15}{\day}$}
    \label{app:alt_fit}
    In this section, we present an alternative joint fit, in which we consider the \SI{15.0}{\day} period to be the signal underlying the aliases discussed in \autoref{subsubsec:aliasing}. The priors are identical to the joint fit in \autoref{subsec:joint_modelling}, except the prior for the period of the second component, which was set uniform between \SIrange{14.71}{15.48}{\day} according to its peak width in the GLS periodogram. The posterior for the transiting planet, the \SI{15.0}{\day} signal and the GP for this fit are shown in \autoref{tab:alt_posterior}, and the resulting planetary parameters for GJ~3929~b are in \autoref{tab:alt_pparam}. We found no significant deviation and almost identical uncertainties {for the transiting planet} compared to the values obtained from considering the {\fourteenday} period. {However, the mass of the \fifteenday candidate is noticeably smaller and more uncertain compared to the fit considering the \fourteenday period. Nevertheless}, the higher planetary mass of GJ~3929~b derived from the \fifteenday fit, which was already evident in the RV-only fit, in combination with the consistent uncertainties, leads to a significant ($>3\sigma$) mass measurement.

    \begin{table}[!ht]
        \caption{Median posterior parameters from the alternative joint fit for the transiting planet, the \SI{\sim15.0}{\day} signal, and the GP.}
        \label{tab:alt_posterior}
        \begin{tabular}{lcl}
            \hline \hline
            Parameter                                  & Posterior\tablefootmark{(a)}                            & Units                            \\
            \hline
            \noalign{\smallskip}
            \multicolumn{3}{c}{\textit{Stellar density}}                                                                                            \\
            $\rho_\star$                               & {$\num{14.99}^{+\num{0.42}}_{-\num{0.42}}$}             & \si{\gram\per\centi\meter\cubed} \\
            \noalign{\smallskip}
            \multicolumn{3}{c}{\textit{GJ~3929~b}}                                                                                                  \\
            $P_\text{{b}}$                             & $\num{2.6162749}^{+\num{2.8e-06}}_{-\num{2.8e-06}}$     & d                                \\
            $t_{0,\text{{b}}}$\tablefootmark{(b)}      & $\num{2459320.05808}^{+\num{0.00018}}_{-\num{0.00018}}$ & d                                \\
            $r_{1,\text{{b}}}$                         & $\num{0.41}^{+\num{0.045}}_{-\num{0.045}}$              & \dots                            \\
            $r_{2,\text{{b}}}$                         & $\num{0.03349}^{+\num{0.00043}}_{-\num{0.00042}}$       & \dots                            \\
            $K_\text{{b}}$                             & $\num{1.55}^{+\num{0.46}}_{-\num{0.45}}$                & $\mathrm{m\,s^{-1}}$             \\
            \noalign{\smallskip}
            \multicolumn{3}{c}{\textit{15.0 d signal}}                                                                                              \\
            $P_\text{(15.0 d)}$                        & $\num{15.035}^{+\num{0.085}}_{-\num{0.088}}$            & d                                \\
            $t_{0,\text{(15.0 d)}}$\tablefootmark{(b)} & $\num{2459071.35}^{+\num{1.7}}_{-\num{0.88}}$           & d                                \\
            $K_\text{(15.0 d)}$                        & $\num{1.22}^{+\num{0.82}}_{-\num{0.76}}$                & $\mathrm{m\,s^{-1}}$             \\
            \noalign{\smallskip}
            \multicolumn{3}{c}{\textit{GP parameters}}                                                                                              \\
            $P_\text{GP, rv}$                          & $\num{136.9}^{+\num{7.1}}_{-\num{7.2}}$                 & d                                \\
            $\sigma_\text{GP, rv}$                     & $\num{1.65}^{+\num{0.77}}_{-\num{0.63}}$                & $\mathrm{m\,s^{-1}}$             \\
            {$f_\text{GP, rv}$}                        & $\num{0.58}^{+\num{0.19}}_{-\num{0.2}}$                 & \dots                            \\
            {$Q_{0, \text{GP, rv}}$}                   & $\num{5.6}^{+\num{81.4}}_{-\num{5.1}}$                  & \dots                            \\
            {$dQ_\text{GP, rv}$}                       & $\num{83.0}^{+\num{393.0}}_{-\num{72.0}}$               & \dots                            \\
            \hline
        \end{tabular}
        \tablefoot{\tablefoottext{a}{Error bars denote the $68\%$ posterior credibility intervals.}
            \tablefoottext{b}{Barycentric Julian date in the barycentric dynamical time standard.}}
    \end{table}

    \begin{table}[!ht]
        \addtolength{\tabcolsep}{-5pt}
        \caption{Alternative derived planet parameters for GJ~3929~b {and the planet candidate} considering the \SI{15.0}{\day} period.}
        \label{tab:alt_pparam}
        \begin{tabular}{lccl}
            \hline \hline
            Parameter                                   & {Posterior P$_\text{b}$\tablefootmark{(a)}}       & {Posterior P$_\text{(15.0 d)}$\tablefootmark{(a)}} & Units                            \\
            \hline
            \noalign{\smallskip}
            \multicolumn{4}{c}{\textit{Derived transit parameters}}                                                                                                                                 \\
            \noalign{\smallskip}
            $p = R_{\rm p}/R_\star$                     & $\num{0.03349}^{+\num{0.00043}}_{-\num{0.00042}}$ & \dots                                              & \dots                            \\
            $b = (a_{\rm p}/R_\star)\cos i_{\rm p}$     & $\num{0.115}^{+\num{0.067}}_{-\num{0.067}}$       & \dots                                              & \dots                            \\
            $a_{\rm p}/R_\star$                         & $\num{17.57}^{+\num{0.17}}_{-\num{0.17}}$         & \dots                                              & \dots                            \\
            $i_{\rm p}$                                 & $\num{89.63}^{+\num{0.22}}_{-\num{0.23}}$         & \dots                                              & deg                              \\
            \noalign{\smallskip}
            \multicolumn{4}{c}{\textit{Derived physical parameters\tablefootmark{({b})}}}                                                                                                           \\
            \noalign{\smallskip}
            $M_{\rm p}$                                 & $\num{1.53}^{+\num{0.45}}_{-\num{0.44}}$          & \dots                                              & $M_\oplus$                       \\
            $M_{\rm p}\sin i$                           & $\num{1.53}^{+\num{0.45}}_{-\num{0.44}}$          & $\num{2.1}^{+\num{1.5}}_{-\num{1.4}}$              & $M_\oplus$                       \\
            $R_{\rm p}$                                 & $\num{1.151}^{+\num{0.04}}_{-\num{0.039}}$        & \dots                                              & $R_\oplus$                       \\
            $\rho_{\rm p}$                              & $\num{5.5}^{+\num{1.8}}_{-\num{1.7}}$             & \dots                                              & \si{\gram\per\centi\meter\cubed} \\
            $g_{\rm p}$                                 & $\num{11.3}^{+\num{3.5}}_{-\num{3.3}}$            & \dots                                              & \si{\meter\per\second\squared}   \\
            $a_{\rm p}$                                 & $\num{0.02573}^{+\num{0.00086}}_{-\num{0.00085}}$ & $\num{0.0806}^{+\num{0.0012}}_{-\num{0.0012}}$     & \si{\astronomicalunit}           \\
            $T_\textnormal{eq, p}$\tablefootmark{({c})} & $\num{568.4}^{+\num{9.0}}_{-\num{9.0}}$           & $\num{321.1}^{+\num{7.5}}_{-\num{7.4}}$            & \si{\kelvin}                     \\
            $S$                                         & $\num{17.4}^{+\num{1.3}}_{-\num{1.2}}$            & $\num{1.778}^{+\num{0.057}}_{-\num{0.054}}$        & $S_\oplus$                       \\
            {ESM\tablefootmark{(d)}}                    & $\num{4.8}^{+\num{0.2}}_{-\num{0.2}}$             & \dots                                              & \dots                            \\
            {TSM\tablefootmark{(d)}}                    & $\num{19.8}^{+\num{8.1}}_{-\num{4.6}}$            & \dots                                              & \dots                            \\
            \hline
        \end{tabular}
        \tablefoot{\tablefoottext{a}{Error bars denote the $68\%$ posterior credibility intervals.}
            \tablefoottext{b}{We sample from a normal distribution for the stellar mass, stellar radius and stellar luminosity that is based on the results from \autoref{sec:stellar_prop}.}
            \tablefoottext{c}{Assuming a zero Bond albedo.}
            \tablefoottext{d}{{Emission and transmission} spectroscopy metrics \citep{Kempton.2018}}}
    \end{table}

    \clearpage
    \onecolumn
    \section{{Detailed record of the peaks and aliases visible in the GLS periodograms of the activity indicators}}
    \label{app:activity}
    \begin{table*}[!h]
        \centering
        \caption{Peaks of interest and their appearing aliases in the activity indicator periodograms.}
        \label{tab:aliases}
        \begin{tabular}{rcS[table-format=3.0]S[table-format=3.0]S[table-format=3.0]S[table-format=3.0]S[table-format=3.0]S[table-format=3.0]}
            \hline \hline
            \multirow{2}{*}{Indicator}   &                 & {\multirow{2}{*}{Peak period [d]}} & \multicolumn{4}{c}{Seasonal sampling alias period [d]} & {\multirow{2}{*}{2nd harmonic [d]}}                                                           \\
                                         &                 &                                    & {$m=1$}                                                & {$m=-1$}                            & {$m=2$}           & {$m=-2$}           &                \\

            \hline
            \multirow{2}{*}{CRX}         & \ldelim\{{2}{*} & {\bf \, 70}                        & 56                                                     & 92{$^*$}                            & 47{$^\dagger$}    & 134{$^*$}          & {\dots}        \\
                                         &                 & 128                                & 89                                                     & 228{$^\dagger$}                     & 68{$^*$}          & {\dots}            & {\dots}        \\ \\
            \multirow{2}{*}{H $\alpha$}  & \ldelim\{{2}{*} & { \bf 118}                         & 84                                                     & 198{$^*$}                           & 65                & {\dots}            & {\dots}        \\
                                         &                 & 65                                 & {\dots}                                                & 84                                  & {\dots}           & 117{$^*$}          & {\dots}        \\ \\
            \multirow{2}{*}{Ca II IRT b} & \ldelim\{{2}{*} & 138                                & {\bf \, 94{$^*$}}                                      & 261{$^*$}                           & 71{$^{*\dagger}$} & {\dots}            & 69{$^*$}       \\
                                         &                 & 68{$^\dagger$}                     & {\dots}                                                & {\bf \, 89}                         & {\dots}           & 127{$^*$}          & {\dots}        \\ \\
            \multirow{2}{*}{NaD1}        &                 & 146                                & 97{$^{*\dagger}$}                                      & {\bf 292{$^*$}}                     & {\dots}           & {\dots}            & {\dots}        \\ \\
            \multirow{2}{*}{NaD2}        &                 & 146                                & 97{$^{*\dagger}$}                                      & {\bf 292{$^*$}}                     & {\dots}           & {\dots}            & {\dots}        \\ \\
            \multirow{2}{*}{TiO 7050}    & \ldelim\{{2}{*} & 113{$^\dagger$}                    & 79{$^{*\dagger}$}                                      & 198{$^*$}                           & 61{$^*$}          & {\bf 785{$^*$}}    & {\dots}        \\
                                         &                 & 177                                & 106{$^{*\dagger}$}                                     & {\bf 537{$^*$}}                     & {\dots}           & {\bf 519{$^*$}}    & {\dots}        \\ \\
            TiO 8430                     &                 & {\bf 132}                          & 88                                                     & 264{$^*$}                           & 66{$^\dagger$}    & {\dots}            & 66{$^\dagger$} \\ \\
            \multirow{2}{*}{TiO 8860}    & \ldelim\{{2}{*} & 113{$^\dagger$}                    & 79{$^{*\dagger}$}                                      & 198{$^{*\dagger}$}                  & {\dots}           & 785{$^*$}          & {\dots}        \\
                                         &                 & {\bf 437}                          & 165{$^{*\dagger}$}                                     & 667{$^{*}$}                         & {\dots}           & 189{$^{*\dagger}$} & {\dots}        \\ \\
            \hline
        \end{tabular}
        \tablefoot{{The highest peak in the periodogram is marked in bold. Peaks that are reasonably close to other peaks in the periodogram but do not coincide with the actual centre are denoted with an asterisk, and peaks that are clearly recognisable but below the \SI{10}{\percent} FAP level are flagged with a dagger.}}
    \end{table*}

    \clearpage
\end{appendix}

\end{document}